\documentstyle[12pt]{article}
\input xy
\xyoption{all}

\def\sqr#1#2{{\vcenter{\vbox{\hrule height.#2pt
            \hbox{\vrule width.#2pt height#1pt \kern#1pt
                  \vrule width.#2pt}\hrule height.#2pt}}}}
                                                                                
\def\square
 {\mathop{\mathchoice{\sqr{12}{15}}{\sqr{9}{12}}
{\sqr{6.3}{9}}{\sqr{4.5}{9}}}}

\begin{document}

\hfill hep-th/0606034

\vspace{0.5in}

\begin{center}

{\large\bf Cluster Decomposition, T-duality, and Gerby CFT's}

\vspace{0.2in}

Simeon Hellerman$^1$, Andr\'e Henriques$^2$,
Tony Pantev$^3$, and Eric Sharpe$^4$ \\
(with an appendix by Matt Ando$^5$) \\

\begin{tabular}{cc}
{ \begin{tabular}{c}
$^1$ School of Natural Sciences \\
Institute for Advanced Study \\
Princeton, NJ  08540 
\end{tabular} } &
{ \begin{tabular}{c}
$^2$ 
Mathematisches Institut\\
Westf\"alische Wilhelms-Universit\"at\\
48149 M\"unster, Germany 
\end{tabular} } \\
{ \begin{tabular}{c}
$^3$ Department of Mathematics \\
University of Pennsylvania \\
Philadelphia, PA  19104-6395 
\end{tabular} } &
{ \begin{tabular}{c}
$^4$ Departments of Physics, Mathematics \\
University of Utah \\
Salt Lake City, UT  84112 
\end{tabular} } \\
{ \begin{tabular}{c}
$^5$ Department of Mathematics \\
University of Illinois \\
Urbana, IL  61801
\end{tabular} } & $\,$ 
\end{tabular}

{\tt simeon@ias.edu}, {\tt henrique@math.uni-muenster.de},
{\tt tpantev@math.upenn.edu}, {\tt ersharpe@math.utah.edu},
{\tt mando@math.uiuc.edu} \\

$\,$ \\

\end{center}

In this paper we study CFT's associated to gerbes.
These theories suffer from a lack of cluster decomposition,
but this problem can be resolved:  the CFT's are
the same as CFT's for disconnected targets. 
Such theories also lack cluster decomposition, but in that form,
the lack is manifestly not very problematic.
In particular, we shall see that this matching of CFT's, this duality
between noneffective-gaugings and sigma models on disconnected targets,
is a worldsheet duality related to T-duality.
We perform a wide variety of tests of this claim, ranging from checking
partition functions at arbitrary genus to D-branes to mirror symmetry.
We also discuss a number of applications of these results,
including predictions for quantum cohomology and Gromov-Witten theory
and additional physical understanding of the geometric Langlands program.

\begin{flushleft}
June 2006
\end{flushleft}

\newpage

\tableofcontents

\newpage

\section{Introduction}

In the papers \cite{nr,msx,glsm}, 
the notion of string propagation on stacks was developed.
Stacks are very closely related to spaces, so as strings can propagate
on spaces, it is natural to ask whether strings can also propagate
on stacks.
Briefly, the idea is that nearly every stack has a presentation
of the form $[X/G]$ where $G$ is a not-necessarily-finite,
not-necessarily-effectively-acting group acting on a space $X$,
and to such presentations, one associates a $G$-gauged sigma model
on $X$.  The basic problem is that presentations of this form are
not unique,
and the physics can depend strongly on the proposed dictionary.
For example, a given stack can have presentations
as global quotients by both finite and nonfinite groups;
the former leads immediately to a CFT, whereas the latter will give
a massive non-conformal theory.  The physically-meaningful conjecture 
is that the CFT appearing at the endpoint of renormalization-group flow
is associated to the stack and not a presentation thereof.

Unfortunately, as renormalization group flow cannot be followed explicitly,
to check this claim we must rely on computations in examples and
indirect tests.  To further complicate matters, obvious indirect tests 
have problems.
For example, one of the first things one checks is whether mathematical
deformations of the stack match physical deformations of the 
associated conformal field theory.  Even in very simple examples,
they do not.  This issue and others were addressed in \cite{nr,msx,glsm},
where numerous tests of the conjecture were also presented,
and other consequences derived, such as mirror symmetry for stacks.

For special kinds of stacks known as gerbes, there is an additional
puzzle.  Specifically, the naive massless spectrum computation
contains multiple dimension zero operators, which manifestly
violates one of the fundamental axioms of quantum field theory,
known as cluster decomposition.
On the face of it, this suggests that there is a significant problem
here:  either the massless spectrum computation is wrong, or,
perhaps, there is a fundamental inconsistency in the notion of
strings propagating on gerbes.  Perhaps, for example, renormalization
group flow does not respect these mathematical equivalences,
and physics is presentation-dependent.

There is a loophole, however:  a sigma model describing string propagation
on a disjoint union of several spaces also violates cluster decomposition,
but in the mildest possible way.  We shall argue in this paper that that
is precisely what happens for strings on gerbes:  the CFT of a string
on a gerbe matches the CFT of a string on a disjoint union of spaces
(which may vary in their geometry and background fields), and so both theories are consistent
despite violating cluster decomposition.
We refer to the claim that the CFT of a gerbe matches the CFT of a disjoint
union of spaces as the decomposition
conjecture.

We begin in section~\ref{clusterdecomp} by reviewing how multiple
dimension zero operators violate cluster decomposition.
In section~\ref{reviewgerbes} we review some basics of gerbes, which will
be appearing throughout this paper.
In section~\ref{generaldecomp} we present the general decomposition conjecture.
We claim that the CFT of a string on a gerbe is the same as the CFT
of a string on a disjoint union of (in general distinct) spaces
with $B$ fields; our conjecture specifies the spaces and $B$ fields that
appear.  Much of the rest of this paper is spent verifying that
conjecture.  In section~\ref{exs} we study several examples of gerbes
presented as global quotients by finite noneffectively-acting groups.
In those examples we compute partition functions at arbitrary genus and
massless spectra, and verify the results of the conjecture in each case.
In section~\ref{2dgauge} we discuss gerbes presented as global quotients by
nonfinite noneffectively-acting groups, and how the decomposition conjecture
can be seen directly in the structure of nonperturbative sectors in those
theories.  In section~\ref{Dbranes} we discuss the way in which this decomposition
conjecture can be seen in open strings.  The conjecture makes a prediction
for equivariant K-theory, which we prove.  This conjecture also reflects
a known result concerning sheaf theory on gerbes, which we discuss,
along with other checks of the open string sector of the topological B model.
In section~\ref{mirrorsymm} we use mirror symmetry to give additional checks
of the conjecture.  Mirror symmetry for gerbes (and other stacks) was
worked out in \cite{glsm}; here, we apply it to gerbes and use it to
give another check of the decomposition conjecture.
In section~\ref{noncomm} we give some arguments based on noncommutative-geometry that support
this conjecture.  In section~\ref{dt} we discuss
discrete torsion in noneffective orbifolds and how it modifies the
decomposition conjecture.  In section~\ref{Tdual} we discuss
why we label this decomposition result a ``T-duality.''

In section~\ref{apps} we discuss applications of this decomposition.
One set of applications is to Gromov-Witten theory.
We discuss quantum cohomology, including the analogue of Batyrev's
conjecture for toric stacks outlined in \cite{glsm}, and also discuss
precise predictions for Gromov-Witten theory, looking at two examples
in more detail.  We discuss how these same ideas can be applied to
better understand the behavior of ordinary gauged linear sigma models.
In particular, this should make it clear that the effects we are studying
are not overspecialized, but rather occur relatively commonly.
We also discuss how part of the physical picture of the geometric Langlands
correspondence can be described using these results, and give a physical link between
some mathematical aspects of geometric Langlands not discussed
in \cite{kaped}.  We also speculate on the possible significance of
this decomposition conjecture for the interpretation of nonsupersymmetric
orbifolds.

We conclude in section~\ref{concl}.
In appendices we review some standard results on dilaton shifts and group
theory which we use in this paper.  We also sketch a proof of a result
on equivariant K-theory that checks one of the predictions of our
decomposition conjecture.

\section{Cluster decomposition}    \label{clusterdecomp}

In \cite{nr,msx,glsm}, several examples of theories with physical
fields valued in roots of unity were described.  
These were (mirrors of) theories in which a noneffectively-acting
finite group had been gauged; the physical field valued in roots
of unity was a twist field for the trivially-acting subgroup.

Such theories necessarily fail cluster decomposition,
as we shall now argue.

Let $\Upsilon$ be a twist field for a trivially-acting conjugacy class
in a noneffectively-gauged group.
Because the group element acts trivially, $\Upsilon$ necessarily
has dimension zero and charge zero -- it is nearly the identity,
except that it does not act trivially on fields.
Because of the usual orbifold selection rules,
\begin{displaymath}
< \Upsilon >  \: = \: 0
\end{displaymath}
as discussed in \cite{nr}.

Now, on a noncompact worldsheet,
if cluster decomposition holds, then
\begin{displaymath}
< \Upsilon {\cal O} > \: = \: < \Upsilon >< {\cal O} >
\end{displaymath}
when the two operators are widely separated.
On the other hand, by the usual Ward identities,
since $\Upsilon$ has dimension zero and the CFT is unitary, correlation functions are
independent of the insertion position of $\Upsilon$,
hence regardless of where we insert $\Upsilon$, by cluster
decomposition we should always have
\begin{displaymath}
< \Upsilon {\cal O} > \: = \: < \Upsilon >< {\cal O} >
\end{displaymath}
On the other hand, since $<\Upsilon>=0$ by the orbifold quantum symmetry,
we would be forced to conclude
\begin{displaymath}
< \Upsilon {\cal O} > \: = \: 0
\end{displaymath}
for any other operator ${\cal O}$, which ordinarily we would
interpret as meaning that $\Upsilon = 0$. 

For this reason, one does not usually consider theories with
multiple dimension zero operators.

The theories described in \cite{nr,msx,glsm} have multiple dimension
zero operators, that are not equivalent to the zero operator,
hence those theories cannot satisfy cluster decomposition.

\section{Review of gerbes}   \label{reviewgerbes}

As we shall be discussing gerbes extensively in this paper,
let us take a moment to review some basic properties.

Given a manifold $M$ and a finite group $G$, the easiest way of giving a 
$G$-gerbe over $M$ is to give an action of some group $H$ on some manifold $X$.
We require that the stabilizer of any point $x\in X$ be isomorphic to $G$, 
and the quotient $X/H$ to be isomorphic to $M$.
Such a $G$-gerbe is denoted $[X/H]$.
(In particular, the stacky quotient $[X/H]$ knows about the existence of
trivially-acting $G$, whereas that information is lost when passing
to the ordinary quotient $X/H$.)
However, not every $G$-gerbe on $M$ comes from such a construction (at least with $H$ a finite group).

The general way of getting a gerbe is to have an open cover $\{U_i\}$ of $M$.
One must specify $g_{ijk} \in G$
on triple overlaps and $\varphi_{ij} \in \mbox{Aut}(G)$ on
double overlaps, obeying the constraints
\begin{equation}    \label{firstg}
\varphi_{jk}\circ\varphi_{ij}\:=\:\mbox{Ad}(g_{ijk})\circ\varphi_{ik}
\end{equation}
on triple overlaps and
\begin{equation}\label{secondg}
g_{jk\ell}\,g_{ij\ell}\:=\:\varphi_{k\ell}(g_{ijk})\,g_{ik\ell}.
\end{equation}
on quadruple overlaps.

Given a $G$-gerbe presented as $[X/H]$, 
we may recover such a cocycle as follows.
Pick local sections $s_i:U_i\to X$ of the projection map $X\to M$, and elements $h_{ij}\in H$ 
that transform $s_i$ into $s_j$ on double overlaps.
We then have $\varphi_{ij}=\mbox{Ad}(h_{ij})$ and $g_{ijk}=h_{jk}h_{ij}h_{ik}^{-1}$.

In \cite{hitchin} an analogous description of banded $U(1)$ gerbes
was given; here, we are considering $G$ gerbes for $G$ finite but
not necessarily banded.

Let $\mbox{Out}(G)$ be quotient of $\mbox{Aut}(G)$ by the automorphisms 
of the form $\mbox{Ad}(g)$.
Then the data above defines a principal $\mbox{Out}(G)$-bundle.
Indeed, letting $\phi_{ij}$ be the images of $\varphi_{ij}$ under the map
$\mbox{Aut}(G) \rightarrow \mbox{Out}(G)$, we get from equation~(\ref{firstg})
that
\begin{displaymath}
\phi_{jk}\, \phi_{ij} \: = \: \phi_{ik}.
\end{displaymath}
When that $\mbox{Out}(G)$-bundle is trivial, we shall say that the 
gerbe is banded.
Equivalently, a gerbe is banded when all the $\varphi_{ij}$ can be taken in the group $\mbox{Inn}(G)$ of inner automorphisms of $G$.
If $G$ is abelian, this just means that $\varphi_{ij}=1$, and we then see from (\ref{secondg}) 
that a banded $G$-gerbe corresponds to an element of $H^2(M,G)$.
(Note we are implicitly assuming $G$ is finite.)

Given a $G$-gerbe of the form $[X/H]$, there is always a short exact sequence
\begin{displaymath}
1 \: \longrightarrow \: G \: \longrightarrow \: H \: \longrightarrow \:
K \: \longrightarrow \: 1.
\end{displaymath}
If the induced homomorphism $K\to\mbox{Out}(G)$ is trivial, then the gerbe 
is banded. 
For abelian $G$, this condition then just means that $G$ is central in $H$.

We shall also be interested on gerbes on orbifolds.
Given an effective orbifold $M=[X/K]$ and a short exact sequence as above, 
the non-effective orbifold $[X/H]$ is a $G$-gerbe over $M$.
Again, this does not cover all the examples, and the same discussion 
about cocycles could be carried out in this new context.

\section{General decomposition conjecture}  \label{generaldecomp}

\subsection{General statement}

Suppose we have a $G$-gerbe over $M$, presented as $[X/H]$ where
\begin{displaymath}
1 \: \longrightarrow \: G \: \longrightarrow \: H \: \longrightarrow \:
K \: \longrightarrow \: 1.
\end{displaymath}
Let $\hat{G}$ denote the set of irreps of $G$.
Then we get a natural action of $K$ on $\hat{G}$ given as follows.
Given $k \in K$ and $\rho\in \hat G$, $\rho:G\to U(V)$, we pick a lift $h\in H$ of $k$, and define $k\cdot \rho$ to be the representation $g\mapsto \rho(h^{-1}gh)\in U(V)$.
If $G$ is abelian, this is clearly well defined since any other lift $h'$ will be of the form $h'=g'h$, and thus $h'^{-1}gh'=h^{-1}gh$.
If $G$ is not abelian, this is a little bit more subtle.
Given another lift $h'$ of $k$, we now write it as $h'=hg''$.
The representations $g\mapsto \rho(h^{-1}gh)$ and $g\mapsto \rho(h'^{-1}gh')$ are then intertwined by the operator $\rho(g''):V\to V$,
and thus equal in $\hat G$.

Given a $G$-gerbe presented as $[X/H]$,
define $Y = [( X \times \hat{G} )/K]$,
where $K$ acts on $X$ in its standard effective way,
and acts on $\hat{G}$ in the way described above.
Thus, $Y$ will be a disconnected union of spaces and effective orbifolds.
More precisely, $Y$ has as many connected components as there are orbits of $K$ on $\hat G$.
If $K_1,\ldots ,K_n$ are the stabilizers in $K$ of these various orbits, we then have
\[
Y=[X/K_1]\sqcup\cdots\sqcup [X/K_n].
\]

Furthermore, we can define a natural flat $U(1)$ gerbe on $Y$ as follows.  
For each $\rho \in \hat{G}$, we fix a vector space $V_\rho$ with a representation of $G$ corresponding to $\rho$.
We can then build a vector bundle $E\to X \times \hat{G}$ such that
the restriction to each $X \times \{\rho\}$ is a trivial vector bundle with fiber $V_\rho$. 
Even though dimensions differ across components, this is nevertheless a vector bundle because the fibers are the same on each individual component.
Clearly, $G$ acts on $E$: it acts on $X$ by definition, it acts on $\hat G$ by the trivial action, and it acts on each fiber $V_\rho$ via $\rho$.
But there is an obstruction to making that vector bundle $H$-equivariant. 
That obstruction is exactly the $U(1)$-gerbe we are discussing.  
Namely, the group $H$ acts on the base $X\times \hat G$, but not necessarily on the fibers of the bundle $E$.
If we try to extend the action of $G$ to an action of $H$, we have to make some noncanonical choices.
The resulting isomorphisms will commute only up to phase (using Schur's lemma), and from there we get our $U(1)$-gerbe.

We can be more precise. 
Each component $[X/K_i]$ of $Y$ will come with some discrete torsion coming from a $U(1)$-valued 2-cocycle on $K_i$.
These cocycles are computed as follows. 
Let $\rho_1,\ldots,\rho_n$ be elements of $\hat G$ representing each one of the various $K$-orbits. 
Let $K_i$ be the stabilizer of $\rho_i$ and $H_i$ be its preimage in $H$, so that we have exact sequences
\[
1 \: \longrightarrow \: G \: \longrightarrow \: H_i \: \longrightarrow \:
K_i \: \longrightarrow \: 1.
\]
Pick a section $s:K_i\to H_i$, and let us fix a vector space $V_i$ on which $G$ acts by $\rho_i$.
The fact that $K_i$ fixes $\rho_i$ means that the representations
$g\mapsto \rho_i(g)$ and $g\mapsto \rho_i(h^{-1}gh)$ are equivalent for all $h\in H_i$.
For $h$ of the form $s(k)$, let us pick a map $\tilde\rho_i(k):V_i\to V_i$ that intertwines the above two representations.
Using Schur's lemma, one then checks that the map $g s(k)\mapsto \rho_i(g)\tilde\rho_i(k)$ is a projective representation of $H_i$ on $V_i$.
Moreover, since $G$ has an honest action on $V_i$, the 2-cocycle on $H_i$ actually comes form a 2-cocycle on $K_i$. 
This cocycle can be given explicitely by the formula
\[
c(k,k')\:=\:\tilde\rho_i(k)\,\tilde\rho_i(k')\,\tilde\rho_i(kk')^{-1}\rho_i\Big(s(k)s(k')s(kk')^{-1}\Big)^{-1}.
\]

The decomposition conjecture says that physically, the CFT of
a string on the gerbe is the same as the CFT of a string on $Y$,
and so is the same as a string on copies and covers of $M = [X/K]$,
with variable $B$ fields (using the relation between $B$ fields and discrete
torsion established in \cite{medt,dtrev}).

We have stated this conjecture in a form that appears presentation-dependent,
but in fact the decomposition is independent of the choice of presentation.
A presentation-independent description of $Y$ can be given as follows.
As explained in the previous section, a $G$-gerbe on $M$ induces an $\mbox{Out}(G)$-principal bundle $P\to M$.
Since $\mbox{Out}(G)$ acts on $\hat G$, we can take the associated bundle
\[
Y:=P\times_{\mbox{\small Out}(G)}\hat G.
\]
The gerbe on $Y$ is then the obstruction to finding a vector bundle whose fibers are the given representations of the ineffective stabilizer $G$.

\subsection{Specialization to banded $G$-gerbes}

In the special case of banded gerbes, the general conjecture above
simplifies somewhat.  In particular, the $K$ action on $\hat{G}$
is trivial.
(For example, if $G$ lies in the center of $H$, 
the adjoint action of $K$ on $G$ is trivial, and so is its action on $\hat{G}$.)
In this case, $[(X \times \hat{G})/K]
\cong [X/K] \times \hat{G}$.
Thus, our conjecture becomes the statement
that the CFT of a banded $G$-gerbe over $M$
(where $M$ can be either a manifold or an effective orbifold;
in the notation of the previous section, $M = [X/K]$)
is T-dual to a disjoint union of CFT's, each corresponding
to a CFT on $M$ with a flat $B$ field.
The disjoint union is over irreducible representations of $G$,
and the flat $B$ field is determined by the irreducible representation, 
which defines a map
$Z(G) \rightarrow U(1)$ and hence induces a map from the characteristic class
of the gerbe to flat $B$ fields:
\begin{displaymath}
H^2(M, Z(G)) \: \longrightarrow \: H^2(M, U(1))
\end{displaymath}
In short,
\begin{displaymath}
\mbox{CFT}\,\Big(\mbox{banded $G$-gerbe on $X$} \Big) \:\: 
\stackrel{T \: dual}{\leftarrow \!\!\! - \!\!\!  - \!\!\!  - \!\!\!  - \!\!\! \rightarrow} \:
\coprod_{irreps \: of \: G} \mbox{CFT}\,\Big( X \mbox{with flat $B$ field}
\Big)
\end{displaymath}

\subsection{Evidence}

We shall see several types of evidence for this conjecture in this
paper:
\begin{enumerate}
\item For gerbes presented as global quotients by finite
(noneffectively-acting) groups, in section~\ref{exs}
we describe some completely explicit
calculations of partition functions (at arbitrary genus), massless
spectra, and operator products that support the conjecture.  
For example, for $[T^6/D_4]$, where the ${\bf Z}_2$
acts trivially, a ${\bf Z}_2$ gerbe over $[T^6/{\bf Z}_2\times{\bf Z}_2]$,
we shall see that the partition function of the $[T^6/D_4]$ orbifold
is a sum of the partition functions of the $[T^6/{\bf Z}_2\times{\bf Z}_2]$
orbifolds with and without discrete torsion, and also the massless
spectrum of the $[T^6/D_4]$ orbifold is a sum of the massless spectra
of the $[T^6/{\bf Z}_2\times{\bf Z}_2]$ orbifolds with and without
discrete torsion.  Since discrete torsion is just a choice of
$B$ field on the orbifold, as described in detail in \cite{medt,dtrev},
this example nicely illustrates the conjecture.

\item For gerbes presented as global quotients by nonfinite groups,
we can see this decomposition directly in the structure of the
nonperturbative sectors which distinguish the physical theory on the
gerbe from the physical theory on the underlying space, as discussed
in section~\ref{2dgauge}.

\item At the open string level, D-branes on gerbes decompose according
to the irreducible representation of the band, and there are only
massless
open string states between D-branes in the same irreducible representation.
One mathematical consequence is a statement about K-theory,
whose proof is sketched in an appendix.  Mathematically, this also corresponds 
to a standard result that sheaves on gerbes
decompose into sheaves on copies of the underlying space (and covers
thereof) but twisted by a flat $B$ field
determined as above.  We also discuss how the $A_{\infty}$ algebra of open
string states in the topological B model decomposes, 
and the implications of that statement for the closed string states.
This is discussed in section~\ref{Dbranes}.

\item Mirror symmetry for gerbes, as described in \cite{nr,msx,glsm},
generates Landau-Ginzburg models with discrete-valued fields,
which can be interpreted in terms of a sum over components,
consistent with the decomposition conjecture.
In particular, we shall see in the case of the gerby quintic that
it maps banded gerbes to sums of components in which the complex structure
has shifted slightly between components, so on the original side the
CFT must describe a sum of components in which the complexified
K\"ahler data has been shifted, and using the mirror map we find that
it is only the $B$ field that shifts, not the real K\"ahler form.  
This is discussed in section~\ref{mirrorsymm}.

\item We also have some arguments based in noncommutative geometry
for this same decomposition, see section~\ref{noncomm}.

\item From thinking about general aspects of T-duality we get more arguments
in favor of this conjecture, see section~\ref{Tdual}.

\item From the structure of quantum cohomology rings, we get more evidence
still.  In particular, in \cite{glsm} it was observed how Batyrev's conjecture
for quantum cohomology of toric varieties naturally generalizes to toric
stacks, and we find the conjectured decomposition present implicitly in
that conjecture for quantum cohomology rings.  This is discussed
in
section~\ref{apps}, along with applications of these ideas.

\end{enumerate}

\section{Examples}   \label{exs}

\subsection{Trivial $G$ gerbe}

Consider the trivial $G$ gerbe over a space $X$.
This gerbe can be presented as $[X/G]$ where all of $G$ acts trivially
on $X$.  (Also, since we assume throughout that all stacks are
Deligne-Mumford, $G$ is necessarily finite, though not necessarily 
abelian.)

Let us check that the massless spectrum contains precisely
one copy of the cohomology of $X$ for each irreducible representation
of $G$, as is predicted by our general conjecture.

As in a standard effective orbifold,
the twisted sectors are counted by conjugacy classes of $G$.
Since $G$ acts trivially, each twisted sector contains the same
set of states, namely, a copy of the cohomology of $X$.
Moreover, the number of conjugacy classes is the same as the number
of irreducible representations.
Thus, the massless spectrum contains precisely one copy of the cohomology
of $X$ for each irreducible representation of $G$.

In passing, a beautiful discussion of the twist fields in such orbifolds
can be found in \cite{diFranc}[exercise 10.18].

\subsection{First banded example}   \label{firstd4}

Consider the orbifold $[X/D_4]$, where the ${\bf Z}_2$ center of
$D_4$ acts trivially.
Recall there is a short exact sequence
\begin{displaymath}
1 \: \longrightarrow \: {\bf Z}_2 \: \longrightarrow \:
D_4 \: \longrightarrow \: {\bf Z}_2 \times {\bf Z}_2 \: \longrightarrow \:
1
\end{displaymath}
Let the elements of $D_4$ be denoted by
\begin{displaymath}
\{ 1, z, a, b, az, bz, ab, ba = abz \}
\end{displaymath}
where $a^2 = 1$, $b^2 = z$, and $z$ lies in the center ${\bf Z}_2$.
Let the projection to ${\bf Z}_2 \times {\bf Z}_2$ be denoted with
a bar, so that the elements of ${\bf Z}_2 \times {\bf Z}_2$ are 
\begin{displaymath}
\{ 1, \overline{a}, \overline{b}, \overline{ab} \}
\end{displaymath}
and the projection map sends, for example,
\begin{displaymath}
a, az \: \mapsto \: \overline{a}.
\end{displaymath}

According to our decomposition conjecture, the CFT of this noneffective
orbifold should be the same as two copies of $[X/({\bf Z}_2 \times
{\bf Z}_2)]$, one copy with discrete torsion, the other without.
After all, this is a banded ${\bf Z}_2$ gerbe, so since ${\bf Z}_2$ has
two different irreducible representations, the CFT of the gerbe should
involve two copies of the underlying effective orbifold $[X/({\bf Z}_2 \times
{\bf Z}_2)]$.  Furthermore, since the ${\bf Z}_2$ gerbe is nontrivial,
under those two irreducible representations it induces one vanishing and
one nonvanishing element of $H^2([X/({\bf Z}_2 \times {\bf Z}_2)], U(1))$,
and the nonvanishing element corresponds to discrete torsion in
the ${\bf Z}_2 \times {\bf Z}_2$ orbifold \cite{medt}.
In short, the prediction of the decomposition conjecture is that
\begin{displaymath}
\mbox{CFT}([X/D_4]) \: = \: \mbox{CFT}\left( [X/({\bf Z}_2 \times
{\bf Z}_2)]_{d.t.} \sqcup [X/({\bf Z}_2 \times {\bf Z}_2)]_{w/o \: d.t.} \right).
\end{displaymath}
We shall next verify that prediction by calculating partition functions
(at arbitrary genus) and massless spectra.

Following \cite{nr},
the one-loop partition function of this orbifold is
\begin{displaymath}
Z(D_4) \: = \: \frac{1}{|D_4|} \sum_{ g,h \in D_4; gh=hg} Z_{g,h}
\end{displaymath}
and each of the $Z_{g,h}$ twisted sectors that appears, is the same
as a ${\bf Z}_2 \times {\bf Z}_2$ sector, appearing with multiplicity
$| {\bf Z}_2 |^2 = 4$, except for the 
\begin{displaymath}
{\scriptstyle \overline{a}} \square_{\overline{b}} \, , \: \: \:
{\scriptstyle \overline{a}} \square_{\overline{ab}} \, , \: \: \:
{\scriptstyle \overline{b}} \square_{\overline{ab}}
\end{displaymath}
twisted sectors
of $[X/ ({\bf Z}_2 \times {\bf Z}_2)]$.
Thus, the partition function of the $D_4$ orbifold can be expressed as
\begin{eqnarray*}
Z(D_4) & = & \frac{ | {\bf Z}_2 \times {\bf Z}_2 | }{ | D_4 | }
| {\bf Z}_2|^2  \Big( Z({\bf Z}_2 \times {\bf Z}_2 ) \: - \:
\left(\mbox{some twisted sectors}\right)\! \Big) \\
& = & 2\cdot \Big( Z({\bf Z}_2 \times {\bf Z}_2 ) \: - \:
\left(\mbox{some twisted sectors}\right)\! \Big) 
\end{eqnarray*}

Now, we claim that this partition function is the sum of the one-loop partition
functions for the ${\bf Z}_2 \times {\bf Z}_2$ orbifold with and without
discrete torsion.
Recall from \cite{vafaed} that if we let $\zeta$ denote a generator of the
$k$th roots of unity, and let the boundary conditions in a ${\bf Z}_k \times
{\bf Z}_k$ orbifold be determined by the pair of group elements
\begin{eqnarray*}
T_{\sigma} & = & \left( \zeta^a, \zeta^b \right) \\
T_{\tau} & = &   \left( \zeta^{a'}, \zeta^{b'} \right)
\end{eqnarray*}
then the $k$ possible discrete torsion phases in the ${\bf Z}_k \times
{\bf Z}_k$ orbifold are given by
\begin{displaymath}
\epsilon(T_{\sigma}, T_{\tau}) \: = \: \zeta^{m(a b' - b a')}
\end{displaymath}
for $m \in \{ 0, 1, \cdots, k-1\}$.
In the present case, this means the one-loop twisted sectors
in the ${\bf Z}_2 \times {\bf Z}_2$ orbifold that are multiplied by
a phase are
precisely
\begin{displaymath}
{\scriptstyle \overline{a}} \square_{\overline{b}} \, , \: \: \:
{\scriptstyle \overline{a}} \square_{\overline{ab}} \, , \: \: \:
{\scriptstyle \overline{b}} \square_{\overline{ab}}
\end{displaymath}
the same sectors that were omitted in the $D_4$ orbifold,
and moreover, turning on discrete torsion in the ${\bf Z}_2 \times
{\bf Z}_2$ orbifold means multiplying each of the sectors above by
a sign.

Thus, if we add the one-loop partition functions of the ${\bf Z}_2 \times
{\bf Z}_2$ orbifold with and without discrete torsion,
then it is trivial to see we recover the one-loop partition function
of the $D_4$ orbifold.

It is straightforward to repeat this calculation at two-loops.
Let us follow the notation of \cite[section 4.3.2]{medt}.
Each two-loop sector is defined by four group elements
$(g_1 | h_1 | g_2 | h_2)$ obeying the constraint
\begin{displaymath}
h_1 g_1^{-1} h_1^{-1} g_1 \: = \:
g_2^{-1} h_2 g_2 h_2^{-1}
\end{displaymath}
Let us organize the calculation according to ${\bf Z}_2 \times {\bf Z}_2$
sectors, as before.
For example, the $(\overline{a} | \overline{a} | \overline{a} | \overline{a})$
${\bf Z}_2 \times {\bf Z}_2$ sector is the image of the following
$D_4$ sectors:
\begin{displaymath}
( a, az | a, az | a, az | a, az)
\end{displaymath}
The $(\overline{a}|\overline{a}|\overline{a}|\overline{b})$ 
${\bf Z}_2 \times {\bf Z}_2$ sector, on the other hand, can not arise
from any $D_4$ sector.  More generally,
it is straightforward to check that if the pair $(g_1, h_1)$
defines one of the excluded one-loop ${\bf Z}_2 \times {\bf Z}_2$
sectors, {\it or}
if the pair $(g_2, h_2)$ defines an excluded one-loop sector, but not
both, then the set of four group elements cannot be lifted to a consistent
set of 4 group elements defining a two-loop $D_4$ sector.
However, if either neither pair is excluded, or both are excluded,
then the two-loop ${\bf Z}_2 \times {\bf Z}_2$ sector does lift
to two-loop $D_4$ sectors, in fact lifts to $2^4 = 16$ possible
two-loop $D_4$ sectors.  Thus, we can write
\begin{displaymath}
Z_{2-loop}(D_4) \: = \: 
\frac{| {\bf Z}_2 \times {\bf Z}_2 |^2}{| D_4|^2}| {\bf Z}_2|^4 
\Big( Z_{2-loop}(
{\bf Z}_2 \times {\bf Z}_2) \: - \: \left(\mbox{some sectors}\right)
\!\Big)
\end{displaymath}
Furthermore, we can analyze the effects of discrete torsion as before.
Because the ${\bf Z}_2 \times {\bf Z}_2$ orbifold is abelian and so factorizes,
we can write
\begin{displaymath}
\epsilon( g_1, h_1 ; g_2, h_2 ) \: = \:
\epsilon( g_1, h_1) \epsilon(g_2, h_2)
\end{displaymath}
for the two-loop discrete-torsion phase factor, due to factorization.
Thus, two-loop sectors with the property that either $(g_1, h_1)$ 
or $(g_2, h_2)$ at one-loop would have gotten a sign factor, but not
both, will get a sign at two-loops, whereas other sectors remain invariant.
So, summing over two-loop partition functions for ${\bf Z}_2 \times
{\bf Z}_2$ orbifolds with and without discrete torsion will yield
\begin{displaymath}
2\cdot\Big( Z_{2-loop}(
{\bf Z}_2 \times {\bf Z}_2) \: - \: \left(\mbox{some sectors}\right)\!\Big)
\end{displaymath}
where the pattern of omitted sectors is identical.
More generally, it should now be straightforward to see that at
$g$-loops,
\begin{displaymath}
Z_{g-loop}(D_4) \: = \: 
\frac{| {\bf Z}_g \times {\bf Z}_2 |^g}{| D_4|^2}| {\bf Z}_2|^{2g} 
\Big( Z_{g-loop}(
{\bf Z}_2 \times {\bf Z}_2) \: - \: \left(\mbox{some sectors}\right)
\!\Big)
\end{displaymath}
where the overall numerical factor is 
\begin{displaymath}
\frac{| {\bf Z}_2 \times {\bf Z}_2 |^g}{| D_4|^2}| {\bf Z}_2|^{2g} 
\: = \: 2^g \: = \: 2\cdot 2^{g-1}
\end{displaymath}
In other words, the $g$-loop partition function can be written
\begin{displaymath}
Z_{g-loop}(D_4) \: = \: (\sqrt{2})^{2g-2} \cdot Z_{g-loop}\left(
\mbox{two ${\bf Z}_2 \times {\bf Z}_2$ orbifolds, w/ and w/o discrete torsion}
\right)
\end{displaymath}
The overall factor of $(\sqrt{2})^{2g-2}$ can be absorbed into
a dilaton shift (see appendix~\ref{dilatonshift}), and so is of no
physical significance.
So, the partition function of the $D_4$ orbifold matches that
of two ${\bf Z}_2 \times {\bf Z}_2$ orbifolds, with and without
discrete torsion, at arbitrary genus,
exactly as predicted by the decomposition conjecture.

Next let us compute the massless spectrum in a specific example.
Take $X = T^6$, with the same ${\bf Z}_2 \times {\bf Z}_2$ action
described in \cite{vafaed}.
The group $D_4$ has five conjugacy classes, namely,
\begin{displaymath}
\{1\}, \{z\}, \{a, az\}, \{b, bz\}, \{ab, ba\}
\end{displaymath}
in the notation of the appendix, hence there are five sectors in
the space of massless states.
From \cite{vafaed}, the $D_4$-invariant (same as 
${\bf Z}_2 \times {\bf Z}_2$-invariant) massless states in the untwisted
sector of the Hilbert space are given by
\begin{displaymath}
\begin{array}{ccccccc}
 & & & 1 & & & \\
 & & 0 & & 0 & & \\
 & 0 & & 3 & & 0 & \\
1 & & 3 & & 3 & & 1 \\
 & 0 & & 3 & & 0 & \\
 & & 0 & & 0 & & \\
 & & & 1 & & & \end{array}
\end{displaymath}
and from the $\{z\}$ sector we get another copy of the same set of states,
since $z$ acts trivially.
In each of the three remaining twisted sectors we get sixteen copies of
that part of
\begin{displaymath}
\begin{array}{ccccccc}
 & & & 0 & & & \\
 & & 0 & & 0 & & \\
 & 0 & & 1 & & 0 & \\
0 & & 1 & & 1 & & 0 \\
 & 0 & & 1 & & 0 & \\
 & & 0 & & 0 & & \\
 & & & 0 & & & \end{array}
\end{displaymath}
which is invariant under the centralizer of the a representative of
the conjugacy class defining the twisted sector.
However, the centralizer of each of $a$, $b$, $ab$ consists only of $1$, $z$,
the group element itself, and the other element of its conjugacy class,
but from the analysis of \cite{vafaed} it is clear that each of the
states above is invariant under all of the corresponding centralizer.
Thus, in the noneffective $D_4$ orbifold, we have a grand total of
\begin{displaymath}
\begin{array}{ccccccc}
 & & & 2 & & & \\
 & & 0 & & 0 & & \\
 & 0 & & 54 & & 0 & \\
2 & & 54 & & 54 & & 2 \\
 & 0 & & 54 & & 0 & \\
 & & 0 & & 0 & & \\
 & & & 2 & & & \end{array}
\end{displaymath}
massless states.

Next let us compare to the massless spectrum of a CFT given by
the sum of the CFT's corresponding to the ${\bf Z}_2 \times {\bf Z}_2$
orbifold with and without discrete torsion.
From \cite{vafaed} the massless spectrum of the original
${\bf Z}_2 \times {\bf Z}_2$ orbifold, without discrete torsion, 
is given by
\begin{displaymath}
\begin{array}{ccccccc}
 & & & 1 & & & \\
 & & 0 & & 0 & & \\
 & 0 & & 51 & & 0 & \\
1 & & 3 & & 3 & & 1 \\
 & 0 & & 51 & & 0 & \\
 & & 0 & & 0 & & \\
 & & & 1 & & & \end{array}
\end{displaymath}
and the massless spectrum of the ${\bf Z}_2 \times {\bf Z}_2$ orbifold with
discrete torsion turned on is given by
\begin{displaymath}
\begin{array}{ccccccc}
 & & & 1 & & & \\
 & & 0 & & 0 & & \\
 & 0 & & 3 & & 0 & \\
1 & & 51 & & 51 & & 1 \\
 & 0 & & 3 & & 0 & \\
 & & 0 & & 0 & & \\
 & & & 1 & & & \end{array}
\end{displaymath}
so the disjoint sum of these two CFT's would have a grand total of
\begin{displaymath}
\begin{array}{ccccccc}
 & & & 2 & & & \\
 & & 0 & & 0 & & \\
 & 0 & & 54 & & 0 & \\
2 & & 54 & & 54 & & 2 \\
 & 0 & & 54 & & 0 & \\
 & & 0 & & 0 & & \\
 & & & 2 & & & \end{array}
\end{displaymath}
massless states.
Note that this exactly matches the result of the massless spectrum
computation in the $D_4$ orbifold, hence confirming our conjecture.

One can also extract from this computation that the decomposition
of the noneffective orbifold into a sum of theories is not precisely
the same thing as a decomposition by twisted sectors.
For example, states in the $\{a,az\}$ twisted sector contribute to
both the ${\bf Z}_2\times{\bf Z}_2$ orbifolds, with and without discrete
torsion.

\subsection{Closely related example}

Another ${\bf Z}_2$ gerbe over the orbifold $[T^6/{\bf Z}_2 \times {\bf Z}_2]$
is given by $[T^6/{\bf H}]$
where ${\bf H}$ is the nonabelian eight-element finite
group, whose elements are the unit quaternions:
\begin{displaymath}
{\bf H} \: = \: \{ \pm 1, \pm i, \pm j, \pm k \}
\end{displaymath}
This gerbe is distinct from the ${\bf Z}_2$ gerbe over $[T^6/{\bf Z}_2 \times
{\bf Z}_2]$ given by $[T^6/D_4]$, though it is closely related.

In particular, although the gerbe is distinct, the prediction of
the decomposition conjecture is the same in this case as for
$[T^6/D_4]$, namely,
\begin{displaymath}
\mbox{CFT}\,([T^6/{\bf H}]) \: = \: \mbox{CFT}\left(
[T^6/{\bf Z}_2 \times {\bf Z}_2]_{d.t.} \sqcup
[T^6/{\bf Z}_2 \times {\bf Z}_2]_{w/o \: d.t.} \right)
\end{displaymath}
The derivation from the decomposition conjecture is
 identical to the last example, so we omit it here.

In this subsection, we shall check that the one-loop partition function
and massless spectrum of the noneffective ${\bf H}$ orbifold
match that of the $D_4$ orbifold, which in turn describes the CFT
of a sum of ${\bf Z}_2 \times {\bf Z}_2$ orbifolds with and without
discrete torsion.  Thus, although the gerbes are distinct,
their physical descriptions are identical.

The group ${\bf H}$ fits into a short exact sequence just like the
one for $D_4$:
\begin{displaymath}
1 \: \longrightarrow \: {\bf Z}_2 \: \longrightarrow \: {\bf H}
\: \longrightarrow \: {\bf Z}_2 \times {\bf Z}_2 \: \longrightarrow \: 1
\end{displaymath}

First, let us consider the one-loop partition function of the
${\bf H}$ orbifold.  Just as in the $D_4$ orbifold, each
${\bf Z}_2 \times {\bf Z}_2$ sector that appears, appears with
multiplicity $| {\bf Z}_2|^2 = 4$, reflecting the fact that
there is a two-fold ambiguity in lifts of elements of
${\bf Z}_2 \times {\bf Z}_2$, and that the ambiguity is by elements of
the center of ${\bf H}$.

Thus, schematically the one-loop partition function of the ${\bf H}$
orbifold is given by
\begin{eqnarray*}
Z_{1-loop}({\bf H}) & = & \frac{ | {\bf Z}_2 \times {\bf Z}_2 |}{| {\bf H}|}
| {\bf Z}_2 |^2 \Big( Z_{1-loop}\left({\bf Z}_2 \times {\bf Z}_2 \right)
\: - \: (\mbox{some sectors})\Big) \\
& = & 2\cdot \Big( Z_{1-loop}\left({\bf Z}_2 \times {\bf Z}_2 \right)
\: - \: (\mbox{some sectors})\Big) 
\end{eqnarray*}
Denote the elements of ${\bf Z}_2 \times {\bf Z}_2$ by
$\{ 1, \overline{\imath}, \overline{\jmath}, \overline{k} \}$.
Then, since for example $\pm i \in {\bf H}$ can only commute with
$\pm 1$ and $\pm i$, we see that the following ${\bf Z}_2 \times
{\bf Z}_2$ twisted sectors are omitted from the one-loop partition
function of the ${\bf H}$ orbifold:
\begin{displaymath}
{\scriptsize \overline{\imath}} \square_{\overline{\jmath}}, \: \:
{\scriptsize \overline{\imath}} \square_{\overline{k}}, \: \:
{\scriptsize \overline{\jmath}} \square_{\overline{k}}
\end{displaymath}
These are the same one-loop sectors that were omitted in the
$D_4$ case, and the multiplicative factors are the same between here
and the $D_4$ case, hence we see that
\begin{displaymath}
Z_{1-loop}({\bf H}) \: = \: Z_{1-loop}(D_4)
\end{displaymath}
the one-loop partition function of the ${\bf H}$ orbifold matches
that of the $D_4$ orbifold, and hence can also be written as a sum
of the one-loop partition functions of the ${\bf Z}_2 \times
{\bf Z}_2$ orbifolds with and without discrete torsion.

Next, let us check the massless spectrum of the noneffective ${\bf H}$ orbifold,
and compare it to the noneffective $D_4$ orbifold.
The group ${\bf H}$ has five conjugacy classes, given by
\begin{displaymath}
\{ 1 \}, \: \:
\{ -1 \}, \: \:
\{ \pm i \}, \: \:
\{ \pm j \}, \: \:
\{ \pm k \}
\end{displaymath}
The contribution to the massless spectrum of the noneffective
${\bf H}$ orbifold from the conjugacy classes above matches the
contribution to the massless spectrum of the $D_4$ orbifold from the
\begin{displaymath}
\{ 1 \}, \: \:
\{ z \}, \: \:
\{ a, az \}, \: \:
\{ b, bz \}, \: \:
\{ ab, ba \}
\end{displaymath}
conjugacy classes of $D_4$, respectively.
Hence the massless spectrum of the noneffective ${\bf H}$ orbifold
matches that of the noneffective $D_4$ orbifold,
and in particular, is a sum of the massless spectra of the 
${\bf Z}_2 \times {\bf Z}_2$ orbifolds with and without discrete torsion.

\subsection{First nonbanded example}

\subsubsection{Partition function analysis}

Let ${\bf H}$ denote the eight-element group of quaternions, {\it i.e.},
\begin{displaymath}
{\bf H} \: = \: \{ \pm 1, \pm i, \pm j, \pm k \}
\end{displaymath}
Consider $[X/{\bf H}]$ where the subgroup $\langle i \rangle  \cong {\bf Z}_4$ acts
trivially.  There is a short exact sequence
\begin{displaymath}
1 \: \longrightarrow \: \langle i \rangle  \: \longrightarrow \: {\bf H} \:
\longrightarrow \: {\bf Z}_2 \: \longrightarrow \: 1
\end{displaymath}
but note that this is not a central extension -- $\langle i \rangle $ does not lie
in the center of ${\bf H}$.

First, let us work out the prediction of our decomposition conjecture
for this case.  The group $K = {\bf Z}_2$ acts on
$G = \langle i \rangle  = {\bf Z}_4$ by conjugation, and for the purposes of computing
actions on irreducible representations, we can just consider 
conjugation by $j$ representing a coset in  ${\bf H}/\langle i \rangle $.
Under such conjugation, $\pm 1 \mapsto \pm 1$ and $\pm i \mapsto
\mp i$.  Now, $\langle i \rangle $ has four irreducible representations, and under
this action of $K$, two of the irreducible representations are invariant,
and two are exchanged.  The two invariant irreducible representations
correspond to two copies of $[X/{\bf Z}_2]$ in the decomposition,
whereas the two remaining ones intertwine $[X/{\bf Z}_2]$ to get
a two-fold cover, which will be $X$.  (After all, that component is given
by $[(X \times {\bf Z}_2)/{\bf Z}_2]$ and since the ${\bf Z}_2$ acts
nontrivially on the ${\bf Z}_2$, this is just $X$.)
Furthermore, one can show
that the $B$ fields on each component will be trivial.
Thus, the decomposition conjecture predicts 
\begin{displaymath}
\mbox{CFT}( [X/{\bf H}]) \: = \: \mbox{CFT}\Big(
[X/{\bf Z}_2] \sqcup [X/{\bf Z}_2] \sqcup X \Big).
\end{displaymath}
We shall check this statement at the level of partition functions
and operators.

This example was discussed in section~2.0.4 of \cite{nr},
and we argued there that in terms of ${\bf Z}_2$ path integral twisted
sectors, the one-loop partition function has the form
\begin{displaymath}
Z([X/{\bf H}]) \: = \: \frac{1}{| {\bf H} |}
\left( (16) {\scriptstyle 1} \square_1 \: + \:
(8) {\scriptstyle 1} \square_{\xi} \: + \: (8) {\scriptstyle \xi}
\square_{\xi} \right)
\end{displaymath}
where $\xi$ denotes the generator of the effectively-acting ${\bf Z}_2$.
This can be written as
\begin{displaymath}
Z([X/{\bf H}]) \: = \: (2) Z([X/{\bf Z}_2]) \: + \:
Z(X) 
\end{displaymath}
from which we conclude that the CFT of a string on this nonbanded gerbe
is equivalent to the CFT of a string on the disjoint union of two
copies of the orbifold $[X/{\bf Z}_2]$ and one copy of $X$.

Let us also work through a two-loop computation, following
the conventions of \cite[section 4.3.2]{medt}.
Each two-loop sector is defined by four group elements
$(g_1 | h_1 | g_2 | h_2)$ obeying the constraint
\begin{displaymath}
h_1 g_1^{-1} h_1^{-1} g_1 \: = \:
g_2^{-1} h_2 g_2 h_2^{-1}
\end{displaymath}
Let us organize the calculation according to ${\bf Z}_2$ twisted sectors.
For example, the $(1|1|1|1)$ two-loop sector in the ${\bf Z}_2$ orbifold
comes from the
\begin{displaymath}
( \pm 1, \pm i | \pm 1, \pm i | \pm 1, \pm i | \pm 1, \pm i)
\end{displaymath}
two-loop sectors in the noneffective ${\bf H}$ orbifold.
The $(1 | 1| 1| \xi)$ two-loop sector in the ${\bf Z}_2$ orbifold
(where $\xi$ generates ${\bf Z}_2$) comes from the
\begin{displaymath}
( \pm 1, \pm i | \pm 1, \pm i | \pm 1 | \pm j, \pm k ) \\
\end{displaymath}
two-loop sectors in the ${\bf H}$ orbifold.
However, because the group constraint is not satisfied
(because there is no corresponding principal ${\bf H}$-bundle),
there are no
\begin{displaymath}
( \pm 1, \pm i | \pm 1, \pm i | \pm i | \pm j, \pm k)
\end{displaymath}
sectors.
Similarly, the $(1 | 1 | \xi | \xi)$ ${\bf Z}_2$ sectors arise from the
\begin{displaymath}
\begin{array}{c}
( \pm 1, \pm i | \pm 1, \pm i | \pm j | \pm j )\\
( \pm 1, \pm i | \pm 1, \pm i | \pm k | \pm k)
\end{array}\end{displaymath}
${\bf H}$ sectors, the $(1 | \xi | 1 | \xi)$ ${\bf Z}_2$ sectors arise from the
\begin{displaymath}
\begin{array}{c}
( \pm 1 | \pm j, \pm k | \pm 1 | \pm j, \pm k ) \\
( \pm i | \pm j, \pm k | \pm i | \pm j, \pm k)
\end{array}
\end{displaymath}
${\bf H}$ sectors, and the $(1|\xi|\xi|\xi)$ ${\bf Z}_2$ sectors arise from the
\begin{displaymath}
\begin{array}{c}
( \pm 1 | \pm j, \pm k | \pm j | \pm j ) \\ 
( \pm 1 | \pm j, \pm k | \pm k | \pm k ) \\
( \pm i | \pm j, \pm k | \pm j | \pm k ) \\
( \pm i | \pm j, \pm k | \pm k | \pm j )
\end{array}
\end{displaymath}
${\bf H}$ sectors.  We shall not list other cases here, but the analysis
should be clear to the reader.
As a result, we can write
\begin{displaymath}
Z_{2-loop}({\bf H}) \:  = \: \frac{ 1 }{ | {\bf H} |^2 }
\left( 4^4 (\mbox{untwisted sector}) \: + \: 2 \cdot 4^3 (\mbox{all other
sectors}) \right) 
\end{displaymath}
It should now be straightforward to check that more generally,
\begin{eqnarray*}
Z_{g-loop}([X/{\bf H}]) & = & \frac{1}{ | {\bf H} |^g }
\Big(4^{2g}(\mbox{untwisted sector}) \: + \: 2 \cdot 4^{2g-1}(\mbox{all other sectors}) \Big) \\
& = & 
 \frac{ 4^{2g-1} }{8^g} \Big( 2 Z_g(X) \: + \: 2 ( \mbox{sum over all sectors}) \Big) \\
& = & 
 2^{g-2} \Big( 2 Z_g(X) \: + \: 2 | {\bf Z}_2 |^g Z_g([X/{\bf Z}_2]) \Big) \\
& = & 
 (\sqrt{2})^{2g-2} Z_g(X) \: + \: 2^{2g-2} \cdot 2 Z_g([X/{\bf Z}_2])
\end{eqnarray*}
We can get rid of numerical factors of the form $N^{2g-2}$ via
dilaton shifts (see appendix~\ref{dilatonshift}), 
involving giving a vev to the dilaton -- here, since
there are different types of components, one can expect different
dilaton shifts on each component, corresponding to giving the dilaton
a locally-constant vev.
Modulo such dilaton shifts, we see that again the partition function of the
noneffective $[X/{\bf H}]$ orbifold is the same as that of the
disjoint union of two copies of effective $[X/{\bf Z}_2]$ orbifolds
and one copy of $X$, exactly as predicted by the general decomposition
conjecture.

\subsubsection{Operator analysis}

In this subsection we shall analyze the operators in the special
case of the CFT of a bosonic string in the orbifold $[{\bf R}/{\bf H}]$
where the effectively-acting ${\bf Z}_2$ acts on ${\bf R} = X$ by
sign flips.

By the usual logic, the twisted states correspond
to conjugacy classes of group elements.
There are five conjugacy classes: $[1]$, [$-1]$, $[i]$, $[j]$, and $[k]$.
We will soon see that the `good' twist operators -- the
ones which correspond to states in a single universe -- will
actually correspond to certain linear combinations of those
conjugacy classes.

We shall use the notation $\tau_g$ to denote the lowest weight
state in a given twisted sector.  This state is always unique.
In the sectors $\pm 1, \pm i$ with trivially acting twist,
the ground state has weight zero.

The gauge invariant twisted ground states are then the sums of
$\tau_g$
over conjugacy classes.

So we have five gauge invariant twisted ground
states:

\begin{displaymath}
\begin{array}{c}
\sigma_{[1]} \: \equiv \: \tau_1 \\
\sigma_{[-1]} \: \equiv \: \tau_{-1} \\
\sigma_{[i]} \: \equiv \: \frac{1}{2} \left( \tau_i \: + \: \tau_{-i} \right)\\
\sigma_{[j]} \: \equiv \: \frac{1}{2} \left( \tau_j \: + \: \tau_{-j} \right)\\
\sigma_{[k]} \: \equiv \: \frac{1}{2} \left( \tau_k \: + \: \tau_{-k} \right)
\end{array}
\end{displaymath}

The field $X$ (we are using the same notation for the untwisted real
boson as for the space)
is periodic in the first three states and antiperiodic 
in the last two.  Therefore the first three have weight $(0,0)$
and the last two have weight $(1/16, 1/16)$.

This immediately tells us that:

(A) There must be exactly three disconnected components
(in algebraic language, mutually annihilating summands) of the CFT,
since there are three linearly independent operators of weight zero.

(B) Only two of the three components of the CFT can be
orbifold CFT's, since there are only two linearly independent
twisted ground states.

And therefore

(C) The third component must be not an orbifold CFT but
something else.  A likely candidate would be the CFT on the
unorbifolded line X.  But for that to happen, the states
odd under the group element $k$ would have to be restored
in that factor -- otherwise modular invariance could not hold.
We will see, presently, that the $k$-odd states are indeed
restored.

{\it States, operators, OPE's and projectors}

To understand how all this works, organize the three
weight-zero states into mutually annihilating projectors,
which we call `universe operators', for the obvious
reason that they are actually projection operators onto
the Hilbert space of strings propagating in one of the three
`universes'.

The lowest state in the untwisted sector is the identity, so
we write $\tau_1 = 1$.  The weights of
operators are
\begin{displaymath}
\begin{array}{c}
\Delta(\tau_{\pm 1}) \: = \: \Delta(\tau_{\pm i} ) \: = \: 0 \\
\Delta(\tau_{\pm j}) \: = \: \Delta(\tau_{\pm k} ) \: = \: \frac{1}{16}
\end{array}
\end{displaymath}
Note that a ground state twist operator $\Delta(\tau_g)$ has
weight zero if and only if $g$ acts trivially on
$X$.

The OPE of twist operators is determined as usual
by conformal invariance and the consistency of
boundary conditions.  For appropriately arranged branch
cuts, we have
\begin{displaymath}
\tau_g(z) \tau_h(0) \: \sim \: | z \overline{z} |^{\Delta(\tau_{gh})
\: - \: \Delta(\tau_g) \: - \: \Delta(\tau_h)}
\tau_{gh}\left(\frac{z}{2} \right)
\end{displaymath}
In particular, the twist operators which act trivially on
$X$ have no singularities with respect to each other, and
they form a ring.  For $g, h$ in the kernel $K$ of
the action on $X$, we have:
\begin{displaymath}
\tau_g(0) \tau_h(0) \: = \: \tau_{gh}(0)
\end{displaymath}
The $\tau_{g,h}$
are annihilated by the virasoro
operators $L_{-1} = \partial_z$ and 
$\overline{L}_{-1} = \partial_{\overline{z}}$
for $g,h\in K$, so they
are independent of position.  For instance the
above equation gives
\begin{displaymath}
\tau_g(z_1) \tau_h(z_2) \: = \: \tau_{gh}(z_3)
\end{displaymath}
for any independent choice of $z_i$.  So when
dealing with ineffectively acting twist fields, we
will drop the coordinate labels.

Now, the twist operators we are talking about here are not
gauge invariant in general (except the ones which lie in
the center), so the statements we are making depend on
the location of branch cuts.  Here we are taking the gauge
in which all branch cuts of twist fields lie in the forward
time direction.  But when we combine the ground state twist
fields into gauge invariant combinations (which means conjugation
invariant -- the gauge projectors act on the twist fields
by conjugation, $\hat{g} \cdot \tau_h = \tau_{ghg^{-1}}$),
we get statements which are independent of where we put
the branch cuts:
\begin{displaymath}
\begin{array}{c}
\sigma_{[1]}^2 \: = \: \sigma_{[1]} \\
\sigma_{[1]} \sigma_{[-1]} \: = \:
\sigma_{[-1]} \sigma_{[1]} \: = \:
\sigma_{[-1]} \\
\sigma_{[1]} \sigma_{[i]} \: = \: \sigma_{[i]} \sigma_{[1]} \: = \:
\sigma_{[i]} \\
\sigma_{[-1]}^2 \: = \: \sigma_{[1]} \\
\sigma_{[i]} \sigma_{[-1]} \: = \: \sigma_{[-1]} \sigma_{[i]} \: = \: 0 \\
\sigma_{[i]}^2 \: = \: \frac{1}{2} \left( \sigma_{[1]} \: + \: \sigma_{[-1]}
\right) 
\end{array}
\end{displaymath}
The objects $\sigma_{[1,-1,i]}$
comprise the set of gauge invariant operators
of weight zero in the theory -- we shall call it the `ground ring.'
The other states/operators of the theory are a module
over them.  In particular, all gauge invariant operators
of a given weight form submodules over the ground
ring.  For instance, the other ground state twist fields
$\sigma_{[j,k]}$ satisfy
\begin{displaymath}
\begin{array}{c}
\sigma_{[1]} \cdot \sigma_{[j,k]} \: = \: \sigma_{[j,k]} \\
\sigma_{[-1]} \cdot \sigma_{[j,k]} \: = \: 0 \\
\sigma_{[i]} \cdot \sigma_{[j,k]} \: = \:
\sigma_{[k,j]} 
\end{array}
\end{displaymath}

Now, there is a preferred basis for the ground ring, in
which each basis element acts as a projection operator which
annihilates the other projectors.
Define:
\begin{displaymath}
\begin{array}{c}
U^1 \: \equiv \: \frac{1}{4} \left( \tau_1 \: + \: \tau_{-1} \: + \:
\tau_i \: + \: \tau_{-i} \right) \: = \: \frac{1}{4}\left(
\sigma_{[i]} \: + \: \sigma_{[-1]} \: + \: 2 \sigma_{[i]} \right) \\
U^2 \: \equiv \: \frac{1}{4}\left( \tau_1 \: + \: \tau_{-1} \: - \:
\tau_i \: - \: \tau_{-i} \right) \: = \: \frac{1}{4}\left(
\sigma_{[1]} \: + \: \sigma_{[-1]} \: - \: 2 \sigma_{[i]} \right)\\
U^3 \: \equiv \: \frac{1}{2}\left( \tau_1 \: - \: \tau_{-1} \right) \: = \:
\frac{1}{2} \left( \sigma_{[1]} \: - \: \sigma_{[-1]} \right)
\end{array}
\end{displaymath}
These combinations are the `universe operators'.  They are
clearly gauge invariant, since they are just linear
combinations the three gauge invariant twisted
ground states which have weight zero.  They are
projection operators which satisfy:
\begin{displaymath}
\begin{array}{c}
U^1 U^1 \: = \: U^1 \\
U^2 U^2 \: = \: U^2 \\
U^3 U^3 \: = \: U^3 \\
U^i U^j \: = \: 0, \: \: i \neq j
\end{array}
\end{displaymath}
Note that the universe operators satisfy the
completeness property, too:
\begin{displaymath}
U^1 \: + \: U^2 \: + \: U^3 \: = \: \sigma_{[1]} \: = \: \tau_1 \: = \: 1
\end{displaymath}
Therefore they can be used to decompose all states and operators in
the theory 
into objects which live inside three 
summand CFT's.  We will now see that $U^{1,2}$ project
onto orbifold CFT's $[{\bf R}/{\bf Z}_2]$ and $U^3$ projects
onto a CFT describing the unorbifolded real line ${\bf R}$.

{\it Physically twisted states}

Next, let us look at the two gauge invariant twisted ground states
$\sigma_{[j]}$ and $\sigma_{[k]}$
which are `physically twisted' -- that is, there is an actual
dynamical field $X$ which is antiperiodic in those
states.  They are gauge invariant and have weight
$(1/16,1/16)$.

The first thing to notice is that there are only two of them --
so they can only exist in two of the three universes, a
first clue that one of the three universes will not be
an $[{\bf R}/{\bf Z}_2]$.

Let us see which universes they live in.  

Having constructed our universe operators, it is easy to
organize the physically twisted fields into
eigenstates of the $U^i$.  Define
\begin{eqnarray*}
T^1  & \equiv & \frac{1}{4} \left( \tau_j \: + \: \tau_{-j} \: + \:
\tau_k \: + \: \tau_{-k} \right)\\
& = & \frac{1}{2}\left( \sigma_{[j]} \: + \: \sigma_{[k]} \right) \\
T^2 & \equiv & \frac{1}{4} \left( \tau_j \: + \: \tau_{-j} \: - \:
\tau_k \: - \: \tau_{-k} \right) \\
& = & \frac{1}{2} \left( \sigma_{[j]} \: - \: \sigma_{[k]} \right)
\end{eqnarray*}
It is straightforward to check that these twisted ground states
are simultaneous eigenstates of the three universe operators:
\begin{displaymath}
\begin{array}{c}
U^i T^j \: = \: \delta^{ij} T^j, \: \: i = 1, 2\\
U^3 T^j \: = \: 0
\end{array}
\end{displaymath}
This shows that universe number three does not contain any
physically twisted states.

{\it States and operators in universes 1 and 2}

Universes 1 and 2 are then fairly straightforward
as CFT.

In particular, the operators 
\begin{displaymath}
U^1 \cos pX, \: U^1 \partial X \overline{\partial} X, \:
U^1 \partial X \partial^5 X ( \partial^9 X)^4, \: T^1, \:
( \partial X )^2 T^1, \: \cdots
\end{displaymath}
have the exact same dimensions and satisfy the
exact same OPE's as the corresponding
operators 
\begin{displaymath}
\cos pX, \: \partial X \overline{\partial} X, \:
\partial X \partial^5 X (\partial^9 X)^4, \:
\tau, \: (\partial X)^2 \tau, \: \cdots
\end{displaymath}
in a single copy of the
standard effective orbifold CFT $[{\bf R}/{\bf Z}_2]$.  (In
this notation $t$ is the ${\bf Z}_2$ twist
field which cuts $X$ in the effective orbifold CFT.)
And obviously the same statement applies with $U^1$, $T^1$
replaced with $U^2$, $T^2$ respectively.

This shows in particular
that each of the two summands has a closed
and complete OPE, and a modular invariant partition function.

{\it Oddness in universe 3}

This presents a bit of a puzzle for universe number three.
This universe would appear to be described by a
CFT which contains only even states such as $U^3 \cos pX$
but no odd states such as $U^3 \sin pX$.
Indeed, the latter operator is projected out by the
gauge projection. 

This would not be a problem if universe number 3 contained physically
twisted states -- but it apparently does not.  So we are
in danger of the third summand not being a modular
invariant CFT.  And yet we know that the full partition
function is modular invariant and that
universes 1 and 2 are modular invariant CFT by themselves.
So universe 3 must be modular invariant as well.

The resolution of this puzzle is that universe number three does indeed
contain `physically odd' states.  Given any odd operator
of the unorbifolded X CFT, such as $\sin pX$, multiply it by
the ineffective twist field 
$\rho \equiv \frac{1}{2}( \tau_i - \tau_{-i} )$
which
is {\it odd} under the action of $k$ (remember, group
elements act on twist fields by conjugation).  The
resulting field $\rho \sin pX$ is even under the
action of $k$ and indeed invariant under all elements of
the quaternion group!

Using the properties of $\rho$:
\begin{displaymath}
\begin{array}{c}
\rho^2 \: = \: U^3 \\
U^{1,2} \rho \: = \: 0 \\
U^3 \rho \: = \: \rho
\end{array}
\end{displaymath}
we see that any $\rho$-dressed
odd operator lives in the third universe, and that the OPE
of two odd operators, each dressed with $\rho$, generates 
an even operator, dressed with $U^3$.
So the $\rho$-dressed odd operators made out of $X$'s play the
role of odd operators made out of $X$'s in the unorbifolded
CFT on ${\bf R}$.

\subsection{Second nonbanded example}

Consider the nonabelian group $A_4$ of alternating permutations of four
elements.  There is a short exact sequence
\begin{displaymath}
1 \: \longrightarrow \: {\bf Z}_2 \times {\bf Z}_2 \: \longrightarrow \:
A_4 \: \longrightarrow \: {\bf Z}_3 \: \longrightarrow \: 1
\end{displaymath}
so we can consider the orbifold $[X/A_4]$ where a ${\bf Z}_2 \times
{\bf Z}_2$ subgroup acts trivially.

The analysis of the decomposition conjecture is straightforward.
The ${\bf Z}_3$ fixes the trivial representation of
${\bf Z}_2 \times {\bf Z}_2$ and permutes the other three irreducible representations.
So from the decomposition conjecture we should get one copy
of $[X/{\bf Z}_3]$ plus a three-fold cover of $[X/{\bf Z}_3]$, which can be shown to be
$X$.  Also, the $B$ fields will be trivial in this case.
Thus, the decomposition conjecture predicts
\begin{displaymath}
\mbox{CFT}\left( [X/A_4] \right) \: = \:
\mbox{CFT}\Big( [X/{\bf Z}_3] \sqcup X \Big)
\end{displaymath}

This example was discussed in section~2.0.5 of \cite{nr},
and we argued there that in terms of ${\bf Z}_3$ path integral twisted
sectors, the one-loop partition function has the form
\begin{displaymath}
Z([X/A_4]) \: = \: \frac{1}{| A_4 |}\left(
(16) {\scriptstyle 1}\square_1 \: + \: (4) {\scriptstyle 1}\square_{\xi}
\: + \: (4) {\scriptstyle 1} \square_{\xi^2} \: + \:
(4) {\scriptstyle \xi}\square_{\xi} \: + \: \cdots \right)
\end{displaymath}
where $\xi$ denotes the generator of the effectively-acting ${\bf Z}_3$.
This can be written as
\begin{displaymath}
Z([X/{\bf A_4}]) \: = \: Z([X/{\bf Z}_3]) \: + \: Z(X)
\end{displaymath}
from which we conclude that the CFT of a string on this nonbanded gerbe
is equivalent to the CFT of a string on the disjoint union of the
effective orbifold $[X/{\bf Z}_3]$ and $X$.

\subsection{Third nonbanded example}

Next we shall consider a family of nonbanded gerbes,
built using the fact that the dihedral group $D_n$ fits into the
short exact sequence
\begin{displaymath}
1 \: \longrightarrow \: {\bf Z}_n \: \longrightarrow \: D_n \:
\longrightarrow \: {\bf Z}_2 \: \longrightarrow \: 1
\end{displaymath}
(Explicitly, the group $D_n$ is generated by the elements $a$, $b$,
where $a^2=1$, $b^n=1$, and $aba = b^{-1}$, so that the group elements are
$\{ 1, b, b^2, \cdots, b^{n-1}, a, ab, \cdots, ab^{n-1} \}$.)
Consider an orbifold $[X/D_n]$ in which the ${\bf Z}_n$ acts trivially,
so that only the ${\bf Z}_2$ quotient acts effectively.

Let us first explain the prediction of the decomposition conjecture.
If $n$ is odd, then $K = {\bf Z}_2$ will leave one irreducible
representation of ${\bf Z}_n$ invariant and switch the rest in pairs,
so we expect 
\begin{displaymath}
\mbox{CFT}([X/D_n]) \: = \: \mbox{CFT}\left(
[X/{\bf Z}_2]\, \sqcup\! \coprod_1^{(n-1)/2} X \right).
\end{displaymath}
If $n$ is even, then there are two invariant irreducible representations
of ${\bf Z}_n$ and the other irreducible representations are switched
in pairs, so we expect
\begin{displaymath}
\mbox{CFT}([X/D_n]) \: = \: \mbox{CFT}\left(
[X/{\bf Z}_2]\,\sqcup\, [X/{\bf Z}_2]\,\sqcup\!  \coprod_1^{(n-2)/2} X \right).
\end{displaymath}
In each case, the $B$ fields are trivial.

Now, let us check that prediction.
Let $\xi$ denote the generator of the effectively-acting ${\bf Z}_2$.
The 
\begin{displaymath}
{\scriptstyle 1} \square_1
\end{displaymath}
sector of $[X/{\bf Z}_2]$ arises from 
\begin{displaymath}
{\scriptstyle 1, b, \cdots, b^{n-1} }\square_{1, b, \cdots, b^{n-1} }
\end{displaymath}
one-loop twisted sectors in $[X/D_n]$, {\it i.e.} has multiplicity $n^2$.

The 
\begin{displaymath}
{\scriptstyle 1}\square_{\xi}
\end{displaymath}
one-loop twisted sector of $[X/{\bf Z}_2]$ arises from
\begin{displaymath}
{\scriptstyle 1} \square_{a, ab, \cdots ab^{n-1} }
\end{displaymath} 
one-loop twisted sectors in $[X/D_n]$.  In addition, if $n$ is even,
then $a$ commutes with $b^{n/2}$, and so the 
\begin{displaymath}
{\scriptstyle b^{n/2} }\square_{a, ab, \cdots ab^{n-1} }
\end{displaymath}
one-loop twisted sectors in $[X/D_n]$ also contribute.
Thus, this $[X/{\bf Z}_2]$ twisted sector has multiplicity $n$ when
$n$ is odd, and multiplicity $2n$ when $n$ is even.

The 
\begin{displaymath}
{\scriptstyle \xi}\square_{\xi}
\end{displaymath}
one-loop twisted sector of $[X/{\bf Z}_2]$ arises from
\begin{displaymath}
{\scriptstyle ab^i}\square_{ab^i}
\end{displaymath}
one-loop twisted sectors in $[X/D_n]$, for $i \in \{ 0, 1, \cdots n-1\}$.  
In addition, if $n$ is even,
the 
\begin{displaymath}
{\scriptstyle ab^{i + n/2} } \square_{ab^i}
\end{displaymath}
one-loop twisted sectors in $[X/D_n]$ also contribute.
Thus, this $[X/{\bf Z}_2]$ twisted sector has multiplicity $n$ when
$n$ is odd, and multiplicity $2n$ when $n$ is even.

Putting this together, 
the one-loop partition function for the $[X/D_n]$ orbifold can be
written as
\begin{displaymath}
Z([X/D_n]) \: = \: \frac{1}{| D_n |}\left(
(n^2) {\scriptstyle 1}\square_1 \: + \: (n) {\scriptstyle 1} \square_{\xi}
\: + \: (n) {\scriptstyle \xi} \square_{\xi} \right)
\end{displaymath}
when $n$ is odd, and
\begin{displaymath}
Z([X/D_n]) \: = \: \frac{1}{| D_n |}\left(
(n^2) {\scriptstyle 1}\square_1 \: + \: (2n) {\scriptstyle 1} \square_{\xi}
\: + \: (2n) {\scriptstyle \xi} \square_{\xi} \right)
\end{displaymath}
when $n$ is even.
This means that
\begin{displaymath}
Z([X/D_n]) \: = \: Z([X/{\bf Z}_2]) \: + \:
\left( \frac{n}{2} \: - \: \frac{1}{2} \right) Z(X)
\end{displaymath}
for $n$ odd, and
\begin{displaymath}
Z([X/D_n]) \: = \: (2) Z([X/{\bf Z}_2]) \: + \:
\left( \frac{n}{2} \: - \: 1 \right) Z(X)
\end{displaymath}
for $n$ even.

This suggests that for $n$ odd, the CFT of the noneffective
$[X/D_n]$ orbifold is the same as the CFT of the disjoint union of 
$[X/{\bf Z}_2]$ and $(n-1)/2$ copies of $X$,
and for $n$ even, the CFT of the noneffective $[X/D_n]$ orbifold is the
same as the CFT of the disjoint union of two copies of
$[X/{\bf Z}_2]$ and $(n-2)/2$ copies of $X$, exactly as the decomposition
conjecture predicts.

\subsection{General comment on partition functions}

In the last few sections we have studied numerous examples of partition
functions of noneffective orbifolds to give evidence for our decomposition
conjecture for strings on gerbes.  In this section we shall make a brief
general observation concerning such partition functions.

In the case of banded $G$-gerbes, 
when comparing the noneffective orbifold $[X/H]$ to the effective orbifold $[X/K]$,
the coefficient of the untwisted $(1,1)$ sector is always equal to the number of conjugacy classes of the trivially-acting subgroup $G\subset H$.
Indeed, the number of ordered pairs of commuting elements of $G$ is given by
\begin{displaymath}
\sum_g | Z(g) | \: = \: \sum_{[g]} | [g] | \left( \sum_{h \in [g]} |Z(h)| 
\right)
\end{displaymath}
where $Z(g)$ denotes the centralizer of $g$,
$[g]$ denotes a conjugacy class represented by $g$, and the sum over
$[g]$ is a sum over conjugacy classes.
Since the number of elements of any conjugacy class is given by
$|[g]| = |G| / |Z(g)|$,
and the order of the centralizer is constant across elements of the
same conjugacy class,
we see that the number of ordered pairs of commuting elements of $G$
is given by
\begin{displaymath}
|G| C
\end{displaymath}
where $C$ is the number of conjugacy classes of $G$ (equivalently, the
number of irreducible representations of $G$).

Using this result, 
in the case of a banded gerbe (in which the theory decomposes as a 
sum of copies of $[X/K]$ with flat $B$ fields), 
we have that the partition function of the $G$-gerbe can be expanded,
in terms of the $K$-orbifold, as
\begin{eqnarray*}
Z & = & \frac{|K|}{|H|}\, |G|\, C\cdot
\Big( {\scriptsize 1} \square_1 \Big) \: + \: \ldots \\
& = & C\cdot\Big( 
{\scriptsize 1} \square_1\Big) \: + \: \ldots
\end{eqnarray*}
Thus, we see that in general the number of components into which a banded gerbe
should decompose should equal the number of irreducible representations
of the band, a result for which we have generated experimental evidence
in the preceding sections.
(For nonbanded gerbes, the corresponding argument is more complicated
as the theory decomposes into copies not only of $[X/K]$ but also
covers thereof.)

\section{Two-dimensional gauge theoretic aspects}   \label{2dgauge}

So far we have discussed numerous examples of noneffective orbifolds,
but the same results also apply to presentations as global quotients by
nonfinite groups.
Gerbes can be described, for example, by two-dimensional abelian gauge
theories with noneffectively-acting groups (meaning, nonminimal charges).
These were discussed in detail in \cite{glsm}, where for example
a detailed discussion of how these are nonperturbatively distinct
from two-dimensional gauge theories with minimal charges was presented.

The decomposition conjecture of this paper can also be seen
in these gerby abelian gauge theories.
One way is to use mirror symmetry ideas to dualize the theory
to one in which the nonperturbative effects have become classical,
as discussed in \cite{glsm}. This will be reviewed in section~\ref{mirrorsymm}.
Alternately, as we shall discuss here, the effect can be seen
immediately in the structure of the nonperturbative effects
without dualizing.

Consider a $U(1)$ gauge theory in which the charges of all
fields are a multiple of $k$.
In such a theory, as discussed in \cite{glsm},
the nonperturbative effects are equivalent to only summing
over bundles whose $c_1$ is divisible by $k$.
(In four dimensions, this violates cluster decomposition.)
This is equivalent to summing over copies of a $U(1)$ gauge theory
with minimally-charged fields and variable values of the theta angle.
If we let the theta angle in the formal sum of gauge theories
vary through ${\bf Z}_k$, then when we sum over gauge theories,
only sectors in which $c_1$ is a multiple of $k$ will contribute,
the others (weighted by nontrivial roots of unity) will cancel out.

More formally,
\begin{displaymath}
\sum_{n=0}^{\infty} a_{kn}q^{kn} \: = \:
\frac{1}{k} \sum_{m=0}^{k-1}
\left( \sum_{n=0}^{\infty} a_n q^n \zeta^{mn}
\right)
\end{displaymath}
for $\zeta$ a $k$th root of unity, using the identity 
\begin{displaymath}
\sum_{m=0}^{k-1} \zeta^{mn} \: = \: \left\{
\begin{array}{cl}
k & n = 0, \pm k, \pm 2k, \cdots \\
0 & \mbox{else}
\end{array}
\right.
\end{displaymath}

Note that the theta angle is playing a role here closely analogous to
that of discrete torsion in orbifolds -- both multiply nonperturbative
contributions to partition functions by phases.  
In effect, the theta angle is a continuous version of discrete torsion.

Thus, a $U(1)$ gauge theory in which the fields have charges that
are multiples of $k$, is equivalent to a formal sum of gauge theories 
with variable theta angles, realizing our picture for banded gerbes
that the physics of a banded gerbe should be equivalent to a sum
of copies of the underlying space with variable $B$ fields.

\section{D-branes}   \label{Dbranes}

\subsection{Equivariant K-theory}

Since D-brane charges are classified by K-theory, our decomposition
conjecture relating conformal field theories of strings on gerbes to
conformal field theories of strings on spaces has a specific prediction
for equivariant K-theory.  Specifically, consider the quotient
$[X/H]$ where
\begin{displaymath}
1 \: \longrightarrow \: G \: \longrightarrow \: H \: \longrightarrow \: K
\longrightarrow \: 1
\end{displaymath}
and $G$ acts trivially on $X$ (and $H$ need not be finite), then
\begin{center}
the $H$-equivariant K-theory of $X$ \\
should be the same as \\
the twisted $K$-equivariant K-theory of $X\times \hat G$
\end{center}
where the twisting is defined by the $B$ field, as in our decomposition
conjecture.

In appendix~\ref{mandoproof} we sketch a proof of the mathematical prediction
above for equivariant K-theory.

\subsection{Sheaf theory}

It is a standard mathematical result that the category of sheaves on a gerbe
decomposes, in a fashion consistent with our conjecture of this paper.
For example, the category of sheaves on a banded $G$-gerbe decomposes
into subcategories indexed by irreducible representations of $G$,
and Ext groups between distinct components vanish. 
This mathematical fact corresponds to a prediction for D-branes in the
topological B model, which are modeled by coherent sheaves.

To help illustrate the general remarks about decomposition of
sheaves and Ext groups, let us consider an easy example,
involving D0-branes in the trivially-acting ${\bf Z}_n$ orbifold
of ${\bf C}$. 

Although the ${\bf Z}_n$ acts trivially on ${\bf C}$, it
can act nontrivially on Chan-Paton factors, as discussed in
more detail in \cite{nr}.

Let $\rho_i$ denote the $ith$ one-dimensional irreducible representation
of ${\bf Z}_n$, where $\rho_0$ is the trivial representation.

Consider two sets of D0-branes on ${\bf C}$, one set containing
\begin{displaymath}
a_0 \: + \: a_1 \: + \: \cdots \: + \: a_{n-1}
\end{displaymath}
D0-branes, all at the origin, with the $i$th set of branes in
the $\rho_i$ representation of ${\bf Z}_n$,
and the other set containing
\begin{displaymath}
b_0 \: + \: b_1 \: + \: \cdots \: + \: b_{n-1}
\end{displaymath}
D0-branes, all at the origin, with the $i$ set of branes again
in the $\rho_i$ representation of ${\bf Z}_n$.

First, note that this decomposition is precisely the decomposition
by characters of the band described formally earlier.
Although the ${\bf Z}_n$ acts trivially on the space, it
can act nontrivially on the D-branes, and its possible actions can
be classified by characters of ${\bf Z}_n$.

Although the reference \cite{kps} discussed D-branes in
orbifolds by effectively-acting finite groups,
the same analysis applies to noneffectively-acting finite groups.
As in \cite{kps}, open string spectra are calculated by Ext groups
between corresponding sheaves.  In this case, the spectral sequence
of \cite{kps} is trivial, and so if ${\cal E}$, ${\cal F}$ denote
the sheaves over the origin
corresponding to the two sets of D-branes, then
\begin{displaymath}
\mbox{Ext}^p_{ [ {\bf C}/{\bf Z}_n ] }\left( {\cal E}, {\cal F} \right)
\: = \: H^0\left(pt, {\cal E}^{\vee} \otimes {\cal F} \otimes
\Lambda^p {\cal N}_{pt/{\bf C}}\right)^{ {\bf Z}_n }
\end{displaymath}
As in \cite{kps}, the normal bundle ${\cal N}_{pt/{\bf C}} = {\cal O}$ has
a ${\bf Z}_n$-equivariant structure induced by the ${\bf Z}_n$ action
on ${\bf C}$, but since the ${\bf Z}_n$ action on ${\bf C}$ is trivial,
the equivariant structure is defined by the trivial
representation $\rho_0$.

Using the fact that $\rho_i^{\vee} = \rho_{n-i}$, it is straightforward
to calculate
that
\begin{displaymath}
h^0\left(pt, {\cal E}^{\vee} \otimes {\cal F} \right)^{ {\bf Z}_n }
\: = \:
h^0\left(pt, {\cal E}^{\vee} \otimes {\cal F} \otimes {\cal N}_{pt/{\bf C}}
\right)^{ {\bf Z}_n }
\: = \:
a_0 b_0 \: + \: a_1 b_1 \: + \: \cdots \: a_{n-1} b_{n-1}
\end{displaymath}
There are no cross-terms because we take ${\bf Z}_n$-invariants,
and the ${\bf Z}_n$ acts nontrivially on Chan-Paton factors.
As a result,
\begin{displaymath}
\mbox{dim } \mbox{Ext}_{ [ {\bf C}/{\bf Z}_n ] }^p\left( {\cal E},
{\cal F} \right) \: = \:
a_0 b_0 \: + \: a_1 b_1 \: + \: \cdots \: a_{n-1} b_{n-1}
\end{displaymath}
for $p = 0,1$, and vanishes for other $p$.

\subsection{$A_{\infty}$ structures factorize}

We have just explained that Ext groups decompose, in the sense that
any Ext group between sheaves associated to distinct irreducible
representations necessarily vanishes.
This implies that two-point functions in the open string B model
factorize between irreducible representations.

We shall argue here that it also implies that all higher $n$-point
functions also factorize, in the sense that an $n$-point function
will only be nonzero if all of the boundaries are associated to the
same irreducible representation.

To see this, we can apply the methods of \cite{paulsheldon},
who showed\footnote{Strictly speaking, there is an implicit assumption
that boundary renormalization group flow respects the mathematical
notion of quasi-isomorphism, and so a more nearly direct calculation
would still be very interesting, but for the purposes of this paper
we shall gloss over that subtlety.} how to compute 
superpotentials from D-branes in the open string
B model.  

The basic idea is that computation of $n$-point functions amounts
to understanding the structure of an $A_{\infty}$ algebra
describing the disc-level open string field theory.
In the case of the open string B model, there is an easy description
of that $A_{\infty}$ algebra, in terms of \v Cech cocycles valued
in local Hom's, on which the various products can all be easily defined
in terms of compositions.
That particular description is not as useful as one would like,
because one would like to phrase it in terms of multiplications
between Ext group elements, which can be described as the cohomology
(with respect to $m_1$) of $A$.

Now, an $A_{\infty}$ structure on $A$ induces an $A_{\infty}$ structure
on $H^*(A)$, and it is that latter $A_{\infty}$ structure which most
directly computes superpotential terms.

In the case at hand, it is a standard fact that Ext groups between
sheaves on gerbes factorize, and similarly the algebra $A$ of
\cite{paulsheldon} also necessarily factorizes.  After all,
if ${\cal E}$ and ${\cal F}$ are sheaves on the gerbe associated to distinct
components, then ${\it Ext}^{\cdot}( {\cal E}, {\cal F}) = 0$, and so
there is no way to get a nonzero product in $A$.
The method of \cite{paulsheldon} for constructing product structures
on $H^{\cdot}(A)$ involves constructing, level-by-level,
products on $H^{\cdot}(A)$ obeying the $A_{\infty}$ algebra and
latching onto corresponding products in $A$ under the map
$i: H^{\cdot}(A) \rightarrow A$, but if a set of products in $A$ vanishes
identically, then in order to be consistent, the corresponding products in
$H^{\cdot}(A)$ must also vanish.

\subsection{Closed string states in the B model}

The fact that the open string states factorize also implies that the
closed string states factorize.  In the topological B model,
the closed string states can be interpreted as Hochschild cohomology.
However, Hochschild cohomology can be interpreted as deformations
of an $A_{\infty}$ algebra, and in particular the Hochschild cohomology
of closed string B model states can be interpreted as
infinitesimal deformations of the open string $A_{\infty}$ algebra.

Let us briefly recall how this procedure works, at least in general terms.
Let $A$ be an associative algebra over the complex numbers.
The Hochschild cochain complex (with coefficients in $A$)
is the sequence of vector spaces
\begin{displaymath}
C^n(A) \: = \: \mbox{Hom}_{ {\bf C} }\left( A^{\otimes n}, A \right),
\: n=0, 1, 2, \cdots
\end{displaymath}
with differential $\delta: C^n(A) \rightarrow C^{n+1}(A)$ defined by
\begin{eqnarray*}
\left( \delta f \right) (a_1, \cdots, a_{n+1} ) & = &
a_1 f(a_2, \cdots, a_n)  \: + \:
(-)^{n+1} f(a_1, \cdots, a_n) a_{n+1}
\\
& & \: + \:
\sum_{i=1}^n (-)^if(a_1, \cdots, a_{i-1}, a_i a_{i+1}, a_{i+2},
\cdots, a_n)
\end{eqnarray*}
The cohomology of $\delta$ in degree $n$ will be denoted
$HH^n(A,A)$ and called the Hochschild cohomology of $A$
(with coefficients in $A$).

Each 2-cocycle $f(a_1, a_2)$ defined an infinitesimal deformation
of the associative product on $A$.  Specifically, if one defines a new product
by
\begin{displaymath}
a * b \: = \: ab \: + \: t f(a,b)
\end{displaymath}
then the product will be associative to first order in $t$ if and only if
$\delta f = 0$.  Furthermore, $\delta$-coboundaries correspond to 
deformations which lead to isomorphic algebras, so one is only interested
in $\delta$-closed objects modulo $\delta$-exact objects.
More generally, the total Hochschild cohomology $HH^*(A)$ can be understood
as classifying infinitesimal deformations of $A$ in the class
of $A_{\infty}$ algebras, see for example \cite[section 5]{penkschwarz}.

In the present case, the fact that the open string $A_{\infty}$ algebra
factorizes trivially implies that the closed string states also factorize,
giving us another perspective on the decomposition statement for the
closed string B model.

\section{Mirror symmetry}    \label{mirrorsymm}

Let us now examine how our decomposition conjecture
is consistent with mirror symmetry for gerbes and stacks,
as discussed in \cite{nr,msx,glsm}.

Mirror symmetry for toric stacks and complete intersections therein
was discussed in \cite{glsm}.  Specifically, it was discussed
how to build Landau-Ginzburg point `mirrors' to the A model on stacks,
following ideas in \cite{mp,hv}.

For example, the Toda dual to the A model on a ${\bf Z}_k$ gerbe over
${\bf P}^{N-1}$ with characteristic class $-n \mbox{ mod }k \in
H^2({\bf P}^{N-1}, {\bf Z}_k)$ is defined by the superpotential
\begin{displaymath}
W \: = \: \exp(-Y_1) \: + \: \cdots \: + \: \exp(-Y_{N-1}) \: + \:
\Upsilon^n \exp(Y_1 \: + \: \cdots \:  + Y_{N-1})
\end{displaymath}
where $\Upsilon$ is a field valued in $k$th roots of unity, rather
than the complex numbers.  In effect, mirror symmetry mapped
the nonperturbative effects that distinguish the A model on a gerbe
from the A model on the underlying space, to a purely classical
effect, namely a discrete- or character-valued field.
(More formally, mirror symmetry seems to be exchanging the class algebra
of the band with the representation ring.)

Such a discrete-valued field can be re-interpreted in terms of a sum
over components, as our decomposition statement predicts.
In particular, the path integral measure contains a discrete sum,
over values of $\Upsilon$, and that sum can be pulled out of the
path integral:
\begin{displaymath}
\int [D Y_i] \sum_{\Upsilon} \exp(-S(Y_i, \Upsilon)) \: = \:
\sum_{\Upsilon} \int[ D Y_i] \exp(-S(Y_i, \Upsilon))
\end{displaymath}
The right-hand side of the equation above can be interpreted 
as the partition function of a disjoint union of theories, indexed
by the value of $\Upsilon$, in line with the general prediction
of our decomposition conjecture.

Consider the gerby quintic discussed in \cite{glsm},
that is a banded ${\bf Z}_k$ gerbe over the quintic hypersurface
in ${\bf P}^4$, described as a hypersurface in the gerbe over
${\bf P}^4$ with characteristic class $-1 \mbox{ mod }k$.
It was argued there that, so long as $5$ does not divide $k$,
the mirror Landau-Ginzburg point should be a ${\bf Z}_5^3$ orbifold of the 
Landau-Ginzburg model with superpotential
\begin{displaymath}
W \: = \: x_0^5 \: + \: \cdots \: x_4^5 \: + \:
\psi \Upsilon x_0 x_1 x_2 x_3 x_4 
\end{displaymath}
where $x_0, x_1, x_2, x_3, x_4$ are ordinary chiral superfields,
$\Upsilon$ is a field valued in $k$th roots of unity,
and $\psi$ is a complex number, mirror to the complexified K\"ahler
parameter of the original theory.

Having a field valued in roots of unity is equivalent to having
a partition function given by a sum over theories deformed by
roots of unity, so already from this description, we see that
the CFT of the gerby quintic can be described as a sum over theories
in which the complexified K\"ahler parameter has been shifted by
$k$th roots of unity.

Now, let us understand the meaning of that shift more systematically.
Recall from \cite[equ'n 5.9]{candelasetal} that the relation between the
complexified K\"ahler parameter $t$ of the original quintic and
the complex structure parameter $\psi$ of the mirror quintic is
given by
\begin{displaymath}
t \: = \: - \frac{5}{2 \pi i} \left\{
\log(5 \psi) \: - \: \frac{1}{\omega_0(\psi)} \sum_{m=1}^{\infty}
\frac{ (5m)!}{(m!)^5 (5\psi)^{5m}}\left[
\Psi(1 + 5m) \: - \: \Psi(1+m)\right] \right\}
\end{displaymath}
where $2\pi i k t$ is the value of the action evaluated on a rational
curve of degree $k$,

Close to large radius, the mirror map is given by
\begin{displaymath}
t \: \cong \: - \frac{5}{2 \pi i} \log(5 \psi)
\end{displaymath}
from which we see that if we multiply $\psi$ by a $k$th root of
unity, where $5$ does not divide $k$, the result is to shift
the purely real part of $t$, not the imaginary part.
Since $t \cong B + i J$, shifting only the real part of $t$
corresponds to shifting $B$ only, not $J$, exactly our proposed
interpretation.
(Away from large radius, geometric meanings can no longer be
meaningfully assigned.)

\section{Noncommutative geometry}   \label{noncomm}

In this section we shall give a derivation of the full decomposition conjecture based on noncommutative geometry.
We shall write it in the language of local orbifolds, 
this having the advantage of showing that the decomposition is presentation independent.

\subsection {Lie groupoids and $C^*$-algebras}

A local orbifold is often modeled by a Lie groupoid ${\cal G}$ \cite{Moe02}. 
The corresponding non-commutative space is then modeled by the convolution $C^*$-algebra $C^*({\cal G})$.
If the stack $\cal X$ is of the form $[X/H]$, then the groupoid $\cal G$ has $X$ as space of objects, and 
has an arrow $x\to y$ for each group element $g$ sending $x$ to $y$.
The corresponding $C^*$-algebra is then the crossed product of the group $H$ with the algebra of functions on $X$.
A same stack $\cal X$ can however be presented by many different groupoids $\cal G$.
The corresponding $C^*$-algebras then won't be isomorphic but merely Morita-equivalent.
We shall write $nc({\cal X})$ for the above `Morita-equivalence class' and call it the non-commutative space 
associated to $\cal X$.

Given a $U(1)$-gerbe $\tau$ on ${\cal X}$, we can twist the above construction.
Such a gerbe associates a phase $c(g_1,g_2)\in U(1)$ to each pair of composable arrows of ${\cal G}$,
where ${\cal G}$ is some groupoid representing $\cal X$.
This data represents a class in $H^2({\cal X},U(1))$. 
One can then form the twisted algebra $C^*({\cal G},\tau)$ as follows.
Like for $C^*({\cal G})$, the elements of $C^*({\cal G},\tau)$ are complex valued functions on the space of arrows of ${\cal G}$,
but their product is now given by
\[
\big(f_1*f_2\big)(g)=\sum_{g_1 g_2=g}c(g_1,g_2)f_1(g_1)f_2(g_2).
\]
Once again, the $C^*$-algebra $C^*({\cal G},\tau)$ depends on the particular choice of groupoid $\cal G$,
but it's Morita-equivalence class $nc({\cal X},\tau)$ only depends on $\cal X$ and $\tau\in H^2({\cal X},U(1))$.


\subsection{The functor of points philosophy}

We quickly review the formalism of functor of points \cite{LMB00}, 
which allows one to talk about a stack $\cal X$ without having to mention the groupoid $\cal G$, 
(or the action of $H$ on $X$ in the case ${\cal X}\simeq [X/H]$).
This philosophy  says that to give a stack, it's enough to
give its set of points, and the following two additional pieces of data:
\begin{itemize}
\item One should say what it means to have an isomorphism between two points.
\item One should say what it means to have a family of points parametrized by a 
given manifold.
\end{itemize}

As an example, if ${\cal X}= [X/H]$, then a point of $\cal X$ is 
an $H$-equivariant map from $H$ to $X$.
An isomorphism between two points $p$ and $q$ is then given by an element $h\in H$
such that $p(gh)=q(g)$ for all $g\in H$.
And an $M$-parametrized family of points is an $H$-principal bundle over $M$, along with an $H$-equivariant map to $X$.

One can also treat non-commutative spaces via the functor of points point of 
view, even though this is less standard and less mathematically rigorous.
Given a $C^*$-algebra $A$, a {\em point} is an irreducible 
representation of $A$ (not an isomorphism class!). Note
that every irreducible representation has $U(1)$ as automorphism group, 
so non-commutative geometry leads naturally to considering $U(1)$-gerbes.
Unfortunately, irreducible representations do not behave well in families.
Namely, a limit of irreducible representations might be suddenly reducible, 
this is the phenomenon known as ``brane fractionation''. 
We shall decide to ignore this technical subtlety.

Using the functor of points language, 
we now explain how one associates a non-commutative space to a local orbifold $\cal X$.
We just declare a point of $nc({\cal X})$ to be a pair $(p,\rho)$, where $p$ is a point of $\cal X$,
and $\rho:\mbox{Aut}(p)\to U(V)$ is an irreducible representation of the automorphism group of $p$ (= stabilizer group) on some vector space $V$.

$U(1)$-gerbes can also be described in that language.
Namely, a $U(1)$-gerbe over some stack $\cal X$ is a stack $\tau$, equipped with a map $\tau\to \cal X$, satisfying the following properties.
Any two points $\cal L$, $\cal L'$, of $\tau$ sitting over a given point $p$ of $\cal X$ are isomorphic,
and the map $\mbox{Aut}({\cal L})\to \mbox{Aut}(p)$ is a central extension by $U(1)$.
One should think of the objects $\cal L$ as things that want to be one dimensional $\bf C$-vector spaces, but only manage to be so locally in $\cal X$.

Given a $U(1)$ gerbe $\tau$ on a stack $\cal X$, then $nc({\cal X},\tau)$ is given as follows.
A point of $nc({\cal X},\tau)$ is a pair $(p,\rho)$, where $p$ is a point of $\cal X$, 
and $\rho$ is an irreducible twisted representation of $\mbox{Aut}(p)$ on some $V\otimes \cal L$.
Here, $V$ is an ordinary vector space, 
$\cal L$ is a point of the $U(1)$-gerbe sitting over $p$, and symbol $\otimes$ means that we identify the two copies of $U(1)$ 
sitting in $U(V)$ and in $\mbox{Aut}({\cal L})$.
A twisted representation assigns to each element $g\in\mbox{Aut}(p)$ an automorphism $\rho(g)=\alpha(g)\otimes \phi(g)$ of $V\otimes \cal L$
with the property that $\phi(g)\mapsto g$ under the map $\tau\to\cal X$.
We then have the usual condition that $\rho(gg')=\rho(g)\rho(g)$.
It should be noted that a twisted representation on $V\otimes \cal L$ always induces a projective representation $g\mapsto\alpha(g)$ of $\mbox{Aut}(p)$ on $V$.

If the stack $\cal X$ is presented by a groupoid $\cal G$, then $nc({\cal X},\tau)$ is presented by the algebra $C^*({\cal G},\tau)$ introduced in the previous section.

\subsection{Non-effective local orbifolds}

A local orbifold ${\cal X}$ is called non-effective if every point 
$p$ of ${\cal X}$ has a 
non-trivial stabilizer group.
In that case one can single out a subgroup of $\mbox{Aut}_0(p)\subset
\mbox{Aut}(p)$ 
called the ineffective stabilizer group.
It is the kernel of the natural action of $\mbox{Aut}(p)$ on the tangent space $T_p{\cal X}$.
The quotient of ${\cal X}$, where we mod out all stabilizer groups by their 
ineffective subgroups is called ${\cal X}_{\mathit eff}$.
It's again a local orbifold, and ${\cal X}$ sits as a gerbe 
over ${\cal X}_{\mathit eff}$ with 
band $G:=\mbox{Aut}_0(p)$.
In other terms, ${\cal X}$ is a bundle over ${\cal X}_{\mathit eff}$ whose 
fibers are $BG$.
By definition, the gerbe is then banded if that bundle is trivial.

\subsection{Finite group gerbes and covering spaces}\label{sec:fgg}

Let ${\cal X}$ be a local orbifold with ineffective stabilizer $G$, and with 
effective quotient $M:={\cal X}_{\mathit eff}$.

From the fiber sequence $BG\to {\cal X}\to M$
we expect to get a fiber sequence
\[
nc(BG)\to nc({\cal X})\to nc(M).
\]
As a space, $nc(BG)$ looks like a disjoint union of $d$ points, 
one for each irreducible representation of $G$.
So one expects $nc({\cal X})$ to be a $d$-fold cover of $nc(M)$.
This is almost the case, but not quite.
Indeed $nc(BG)$ is not exactly $d$ points, it's rather $d$ copies of $BU(1)$.
So $nc({\cal X})$ is a bundle over $M$ with fiber $\coprod^d BU(1)$.
In other words, $nc({\cal X})$ is a $d$-fold cover of $nc(M)$ but 
twisted by a $U(1)$-gerbe.

The underlying cover is easy to describe:
Let $\hat G:=\pi_0(nc(BG))$ denote the set of isomorphism classes of 
irreducible representations of $G$.
The monodromy representation $\pi_1(M)\to\mbox{Out}(BG)$ induces an 
action of $\pi_1(M)$ on $\hat G$.
From it, we then get a cover 
\[
Y:=\widetilde{M}\times_{\pi_1(M)} \hat G,
\]
where $\widetilde{M}$ denotes the universal cover of $M$.
The local orbifold $Y$ can also be described as follows: 
a point of $Y$ is a 
point $p$ of ${\cal X}$, and an isomorphism class of irreducible
representation of $\mbox{Aut}_0(p)$.
The stack $Y$ comes with a canonical $U(1)$-gerbe $\nu$
whose points are pairs $(p,V)$, where $p$ is a point of ${\cal X}$ 
and $V$ an irreducible representation of $\mbox{Aut}_0(p)$.
The map $\nu\to Y$ then forgets the irreducible representation and 
replaces it by its 
isomorphism class.
Given this gerbe $\nu$, we can state the result:

\vspace{.2cm}
\centerline
{The bundle
$BG\to {\cal X}\to M$
induces a bundle $\hat G\to Y\to M$}
\centerline
{and we have
$nc({\cal X})=nc(Y,\nu)$.}
\vspace{.2cm}

\noindent Clearly, $ Y$ decomposes into as many 
connected components as there are orbits of $\pi_1(M)$ on $\hat G$.
It follows from the above statements that the same holds for $nc(\cal X)$.
So the above argument proves the decomposition conjecture for the part of the CFT that 
can be expressed in terms of $nc(\cal X)$. This includes for example sheaf theory, D-branes, and K-theory.

Note that if ${\cal X}$ came initially with a $U(1)$-gerbe $\tau$, 
then $ Y$
would be replaced by
\begin{displaymath}
Y^\tau:=\widetilde{M}\times_{\pi_1(M)} \hat G^\tau,
\end{displaymath}
where $\hat G^\tau=\pi_0(nc(BG,\tau))$ denotes the set of 
isomorphism classes of projective representations of $G$ with
cocycle fixed by $\tau$.
This space has now as many components as there are orbits of $\pi_1(M)$ on $\hat G^\tau$.

\subsection{Cocycle description}

Recall that in terms of \v Cech cocycles, a $G$-gerbe ${\cal X}\to M$ is given by 
an open cover $\{U_i\}$ of $M$,
functions $\varphi_{i,j}\in Aut(G)$ on double intersections, 
and $g_{ijk}\in G$ on triple intersections, satisfying
\begin{equation}\label{agp}
\begin{array}{c}
\varphi_{jk}\circ\varphi_{ij}=Ad(g_{ijk})\circ\varphi_{ik}\\
g_{jk\ell}\,g_{ij\ell}=\varphi_{k\ell}(g_{ijk})\,g_{ik\ell}.
\end{array}
\end{equation}
The cover $Y\to M$ is then locally $U_i\times \hat G$, and the 
glueing maps on the double intersections are given by
$\varphi_{ij}^*:\hat G\to \hat G$.

We now describe the $U(1)$-gerbe $\nu$ on $Y$ in terms of cocycles.
That gerbe is the obstruction to finding a vector bundle on $Y$ 
whose fibers are the required representations of the ineffective 
stabilizer groups (pulled back to $Y$).
So in order to write down the cocycle, we try to build that vector 
bundle and when the construction does not work any more, we read out our cocycle.

The cover $\{U_i\}$ used in (\ref{agp}) induces a cover of $Y$ whose 
elements are pairs $(U_i,[\rho_i])$, $[\rho_i]\in\hat G$.
Pick representatives $(V_i,\rho_i:G\to U(V_i))$ of the isomorphism classes $[\rho_i]$.  
Over double intersections, we then pick a 
$\varphi_{ij}$-equivariant homomorphisms $f_{ij}:V_i\to V_j$.
This would glue into a vector bundle on $Y$ if we had
\[
f_{jk}\circ f_{ij} = \rho_k(g_{ijk})\circ f_{ik}.
\]
So our cocycle is 
\begin{equation}\label{cfy}
c_{ijk}=\rho_k(g_{ijk})\circ f_{ik}\circ f_{ij}^{-1}\circ f_{jk}^{-1}
\end{equation}
which is a $U(1)$-valued function by Schur's lemma.
Note also that the $f_{ij}$, and hence $c_{ijk}$, are constant functions: this means that $\nu$ is a flat $U(1)$-gerbe.

\subsection{The abelian case}

If $G$ is abelian, we can be more explicit in identifying the 
$U(1)$-gerbe on $Y$.
In that case, all the irreducible representations of $G$ are one 
dimensional, so we may pick our representatives of $\hat G$ to be of 
the form
$({\bf C},\rho:G\to U(1))$.
By Schur's lemma, if $({\bf C},\rho_i)$ is isomorphic to 
$\varphi_{ij}^*({\bf C},\rho_j)$, then $1:{\bf C}\to {\bf C}$ is $\varphi_{ij}$-equivariant.
So we can take all our $f_{ij}$ to be $1$.
Our cocycle formula (\ref{cfy}) then simplifies to
\[
c_{ijk}=\rho_k(g_{ijk}).
\]
In other words,
the $U(1)$-gerbe on $Y$ is the image under the 
representation $\rho:G\to U(1)$
of the $G$-gerbe ${\cal X}$ over $M$.
More precisely,
the band of the $G$-gerbe ${\cal X}\to M$ is a local system on $M$.
The pull back to $Y$ of that local system admits a 
homomorphism to the constant local system $U(1)$.
The $U(1)$-gerbe on $Y$ is then obtained from the 
gerbe ${\cal X}\times_M Y\to Y$ by changing coefficients 
along the above homomorphism.

For arbitrary $G$, we get a similar description of the 
$U(1)$-gerbe on the subspace of $Y$ corresponding to the one dimensional 
representations of $G$.

\subsection{The case when the band is trivial}

Similarly to the abelian case, if the band is trivial, we can simplify 
our cocycle formula (\ref{cfy}).
In that case we may assume that the $\varphi_{ij}$ in (\ref{agp}) are all $1$.
It then follows that all the $g_{ijk}$ are central in $G$.  The 
isomorphism classes of $G$-gerbes with trivial band are thus 
parametrized by $H^2(M,Z(G))$, a theorem of Giraud \cite{Gir71}.

In the construction of our cocycle, we may take all the $f_{ij}$ to 
be identity maps, and once again we get
\[
c_{ijk}=\rho_k(g_{ijk}).
\]
So the cocycle for the
$U(1)$-gerbe on $Y$ is the image under the 
representation $\rho:  Z(G)\to U(1)$
of the cocycle in $H^2(M,Z(G))$ for the $G$-gerbe ${\cal X}\to M$.

\subsection{Massless spectrum}

To compute the cohomology of a space $X$, one often uses a cellular decomposition of $X$ to write down the cochain complex $C^*(X)={\bf C}^{\{\mathit cells\, of\,  X\}}$.
The coboundary operator $C^n(X)\to C^{n+1}(X)$ then records which cells are in the boundary of which other cells.

Given a stack $\cal X$, it's free loop space $L\cal X$ contains a substack $I \cal X$ of loops of length zero, commonly called the inertia stack of $\cal X$.
Let $\underline{I \cal X}$ be the quotient topological space of $I \cal X$.
A point of $\underline{I \cal X}$ is then the same thing as a pair $(p,[g])$, where $p$ is a point of $\cal X$ and $[g]$ is a conjugacy class of $\mbox{Aut}(p)$,
two such pairs $(p,[g])$, $(p',[g'])$ being identified if $p$ and $p'$ are isomorphic and the isomorphism sends $[g]$ to $[g']$.
We shall take the point of view that the massless spectrum of $\cal X$ agrees with the cohomology of $\underline{I \cal X}$ 
(which is the same as the cohomology of $I \cal X$ since we're taking complex coefficients), 
and use it to check the decomposition conjecture.
The complex that computes the massless spectrum of $\cal X$ is then
\[
\bigoplus_{\mathit cells\, of\, X}{\bf C}^{\,\{\mathit conjugacy\, classes\, of\, stabilizer\, group\}}
\]
and the differential is a combination of the usual cellular coboundary, and of the adjoint of the map
${\mathit Conj}(G)\to{\mathit Conj}(H)$ induced by a group homomorphism $G\to H$.

The space $\underline{I \cal X}$ however had the disadvantage that it has too many connected components (even if $\cal X$ is an effective orbifold, 
$\underline{I \cal X}$ is typically disconnected),
and thus the conjectured decomposition is not apparent from the geometry of $\cal X$.
For that purpose, $nc(\cal X)$ is better behaved.
Let $\underline{nc}({\cal X})$ be the topological quotient of $nc(\cal X)$. 
A point of $\underline{nc}({\cal X})$ is then a pair $(p,[\rho])$, where $p$ is a point of $\cal X$ and $[\rho]$ 
is an isomorphism class of irreducible representation of the stabilizer group of $p$,
two such pairs $(p,[\rho])$, $(p',[\rho'])$ being identified if $p$ and $p'$ are isomorphic and the isomorphism sends $[\rho]$ to $[\rho']$.
The space $\underline{nc}({\cal X})$ can also be described as the spectrum of any $C^*$-algebra representing $nc(\cal X)$.

A cell of $\underline{nc}({\cal X})$ then corresponds to a cell of $X$, along with an isomorphism class of irreducible representation of its stabilizer.
The topology of $\underline{nc}({\cal X})$ is typically not Hausdorff, but we can nevertheless write down the complex computing its cohomology:
\[
\bigoplus_{\mathit cells\, of\, X}{\bf C}^{\,\{\mathit iso.\, classes\, of\, irreps\, of\, stabilizer\, group\}}
\]
The differential is induced by the cellular coboundary operator, 
and by the map $Rep(H)\to Rep(G)$ coming from a group homomorphism $G\to H$.
It should be noted that the two above complexes are isomorphism, where the isomorphism
\[
{\bf C}^{\,\{\mathit conjugacy\, classes\}}={\bf C}^{\,\{\mathit iso.\, classes\, of\, irreps\}}
\]
is given by the character table.
This isomorphism is compatible with the differentials, and so we are indeed computing the same cohomology in two different ways.

Now we may use the isomorphism $\underline{nc}({\cal X})=\underline{nc}(Y,\nu)$ and the decomposition of
$Y=\widetilde{M}\times_{\pi_1 (M)}\hat G$ coming from the orbits of $\pi_1(M)$ on $\hat G$.
Clearly, the cohomology of a disjoint union of spaces is the direct sum of the cohomologies,
and so this provides the desired decomposition of the massless spectrum.

\section{Discrete torsion}   \label{dt}

So far, apart from a brief mention in section \ref{sec:fgg}, discrete torsion has appeared solely in the context
of effective orbifolds, as a summand in the decomposition of a string
on a gerbe.  
However, in principle given a noneffective orbifold,
one could imagine turning on discrete torsion in the noneffectively-acting
group.  
In this section we study a few examples where we turn 
on discrete torsion in the noneffectively-acting
group, and examine how
the decomposition conjecture is modified.

\subsection{First (banded) example}

For our first example, consider the noneffective orbifold
$[X/({\bf Z}_2 \times {\bf Z}_2)]$, where the first ${\bf Z}_2$
acts trivially but the second acts effectively.  Let $a$ denote the
generator of the first (trivial) ${\bf Z}_2$, and $b$ the generator
of the second.  Let us also turn on discrete torsion in this model;
since $H^2({\bf Z}_2 \times {\bf Z}_2, U(1)) = {\bf Z}_2$,
there is only one choice.
Let us try breaking up the one-loop twisted sectors in the 
noneffective ${\bf Z}_2 \times
{\bf Z}_2$ orbifold into effective ${\bf Z}_2$ twisted sectors.
The 
\begin{displaymath}
{\scriptstyle 1} \square_1
\end{displaymath}
sector in the effective ${\bf Z}_2$ orbifold comes from any of
\begin{displaymath}
{\scriptstyle 1, a}\square_{1, a}
\end{displaymath}
and so has multiplicity $2^2 = 4$.
The
\begin{displaymath}
{\scriptstyle 1}\square_b
\end{displaymath}
sector in the effective ${\bf Z}_2$ orbifold comes from any of
\begin{displaymath}
{\scriptstyle 1, a}\square_{b, ab}
\end{displaymath}
but although the
\begin{displaymath}
{\scriptstyle 1}\square_{b, ab}
\end{displaymath}
sectors contribute with a positive sign, because of discrete torsion
in the noneffective ${\bf Z}_2 \times {\bf Z}_2$ orbifold, 
the sectors 
\begin{displaymath}
{\scriptstyle a}\square_{b, ab}
\end{displaymath}
contribute with a minus sign, and so cancel out.
Thus, the partition function for the noneffective
${\bf Z}_2 \times {\bf Z}_2$ orbifold contains no net contribution
that appears as a 
\begin{displaymath}
{\scriptstyle 1}\square_b
\end{displaymath}
sector in the effective ${\bf Z}_2$ orbifold.
Similarly, the 
\begin{displaymath}
{\scriptstyle b}\square_b
\end{displaymath}
sector of the effective ${\bf Z}_2$ orbifold comes from
\begin{displaymath}
{\scriptstyle b, ab}\square_{b, ab}
\end{displaymath}
sectors in the ${\bf Z}_2 \times {\bf Z}_2$ orbifold, but the
contribution of the sectors
\begin{displaymath}
{\scriptstyle b}\square_b, \: \: \:
{\scriptstyle ab}\square_{ab}
\end{displaymath}
is canceled by the contribution of the sectors
\begin{displaymath}
{\scriptstyle ab}\square_b, \: \: \:
{\scriptstyle b}\square_{ab}
\end{displaymath}
which because of discrete torsion contribute with a sign.

The final result for the one-loop partition functions has the form
\begin{eqnarray*}
Z_{1-loop}\left( [ X/{\bf Z}_2 \times {\bf Z}_2 ] \right) & = &
\frac{4}{| {\bf Z}_2 \times {\bf Z}_2 |} \, {\scriptstyle 1}\square_1 \\
& = & Z_{1-loop}(X)
\end{eqnarray*}
from which we conclude that the CFT of the noneffective
${\bf Z}_2 \times {\bf Z}_2$ orbifold with discrete torsion is
identical to the CFT of $X$.

\subsection{Second example}

Next, let us consider a noneffective ${\bf Z}_k \times {\bf Z}_k$ orbifold,
where we turn on an element of discrete torsion, $H^2(
{\bf Z}_k \times {\bf Z}_k, U(1)) = {\bf Z}_k$.

Let $g$ denote the generator of the first, trivially-acting,
${\bf Z}_k$, and $h$ denote the generator of the second,
effectively-acting, ${\bf Z}_k$.

Here, there is more than a single value of discrete torsion,
as $H^2({\bf Z}_k \times {\bf Z}_k, U(1)) = {\bf Z}_k$.
Let $\zeta$ denote a generator of $k$th roots of unity, then
discrete torsion assigns to the sector
\begin{displaymath}
{\scriptstyle g^a h^b} \square_{g^{a'}h^{b'}}
\end{displaymath}
the phase factor $\zeta^{m(ab' - ba')}$, where
$m \in \{ 0, \cdots, k-1 \}$ indexes the possible values of discrete torsion.

As before, let us calculate the one-loop partition function of the
noneffective orbifold $[X/{\bf Z}_k \times {\bf Z}_k]$, 
relating it to the partition
function of the effective $[X/{\bf Z}_k]$ orbifold.

The untwisted sector of the effective orbifold arises from the
\begin{displaymath}
{\scriptstyle g^a} \square_{g^{a'}}
\end{displaymath}
sectors of the noneffective ${\bf Z}_k \times {\bf Z}_k$.
Discrete torsion leaves these sectors invariant, so the result is that
the untwisted sector of the effective $[X/{\bf Z}_k]$ orbifold appears
with multiplicity $k^2$.

The 
\begin{displaymath}
{\scriptstyle 1}\square_{h^{b'}}
\end{displaymath}
sector of the $[X/{\bf Z}_k]$ orbifold arises from sectors of the
form
\begin{displaymath}
{\scriptstyle g^a}\square_{g^{a'}h^{b'}}
\end{displaymath}
in the noneffective ${\bf Z}_k \times {\bf Z}_k$ orbifold.
These sectors each pick up phases; the result is that this
sector of the $[X/{\bf Z}_k]$ orbifold arises with a factor
\begin{displaymath}
\sum_{a,a'} \zeta^{m(ab')} \: = \:
k \sum_{a} \left( \zeta^{mb'} \right)^a \: = \:
\left\{ \begin{array}{cl}
0 & mb' \not\equiv 0 \mbox{ mod } k \\
k^2 & mb' \equiv 0 \mbox{ mod } k
\end{array}
\right.
\end{displaymath}

Finally, the
\begin{displaymath}
{\scriptstyle h^b}\square_{h^{b'}}
\end{displaymath}
sectors of the effective $[X/{\bf Z}_k]$ orbifold arise from sectors
of the form
\begin{displaymath}
{\scriptstyle g^a h^b }\square_{g^{a'} h^{b'}}
\end{displaymath}
in the noneffective ${\bf Z}_k \times {\bf Z}_k$ orbifold.
These sectors each pick up phases; the result is that these sectors
of the $[X/{\bf Z}_k]$ orbifold arise with a factor
\begin{eqnarray*}
\sum_{a,a'} \zeta^{m(ab'-ba')} & = &
\left[ \sum_a \left( \zeta^{mb'} \right)^a \right]
\left[ \sum_{a'} \left( \zeta^{-mb} \right)^{a'} \right] \\
& = & \left\{ \begin{array}{cl}
0 & mb' \not\equiv 0 \mbox{ mod } k \mbox{ or } mb \not\equiv 0 \mbox{ mod } k \\
k^2 & mb' \equiv 0 \mbox{ mod }k\mbox{ and } mb \equiv 0 \mbox{ mod } k
\end{array}
\right.
\end{eqnarray*}

When $k$ is prime, the product $mb$ can never be divisible by $k$,
hence the one-loop partition function of the ${\bf Z}_k \times
{\bf Z}_k$ orbifold is given by
\begin{eqnarray*}
Z_{1-loop}([X/{\bf Z}_k \times {\bf Z}_k]) & = &
\frac{k^2}{| {\bf Z}_k \times {\bf Z}_k |}
{\scriptstyle 1}\square_1\\
& = & Z(X)
\end{eqnarray*}
for {\it any} nonzero discrete torsion.

\section{T-duality}   \label{Tdual}

So far in this paper we have described a duality between
strings on gerbes and strings on disjoint unions of spaces.
We believe that this duality should be interpreted as a T-duality,
for reasons we shall explain here.

First, the duality in question is an isomorphism of conformal field
theories, which excludes for example S-duality.

Second, this duality can often be understand as a Fourier-Mukai transform,
another hallmark of T-duality.
This is exactly what happens in \cite{tonyronTdual}, for example.
If you fix an elliptic fibration $X \to B$, and a class
$\alpha \in H^{2}({\cal O}_{X}^{\times})$ then the derived category
of the gerbe on $X$ classified by $\alpha$ is Fourier-Mukai dual to the
derived category of the disconnected union of spaces $Y_{k}$, $k$ - an
integer, where $Y_{k}$ is a deformation of the dual elliptic fibration $X$
corresponding to the class $k\alpha$. 
The kernel of the Fourier-Mukai transform is 
essentially the usual Poincare sheaf.  More precisely, say that we have a
trivial ${\bf Z}_k$ gerbe on $E$. Denote this gerbe by
$X$. Consider the disconnected space $E\times ({\bf Z}_k)^{\vee}$
where $({\bf Z}_k)^{\vee}$ is the group of characters of ${\bf
Z}_k$. Now on the product $X\times (E\times ({\bf Z}_k)^{\vee})$
we have a line bundle $L$ which on each component $X\times (E,\chi)$
is simply the Poincare bundle on $E\times E$ but lifted to the gerbe
as a weight $\chi$ sheaf.

Next, let us briefly outline another, much more handwaved, method
to relate our duality to T-duality.
Begin with an ordinary T-duality relating $S^1$ of radius $R$
to an $S^1$ of radius $1/R$.  Now, gauge $U(1)$ rotations on the
first $S^1$ that rotate the $S^1$ $k$ times instead of once.
Such a gauging should kill all perturbative (momentum) modes,
but should leave $k$ winding modes:  after all, if we quotient by
windings that wrap $k$ times, then a string that wraps only once should
survive.  Mathematically, gauging an $S^1$ by rotations that wind $k$ times
is equivalent to the stack $[\mbox{point}/{\bf Z}_k] = B {\bf Z}_k$.
Now, consider the T-dual gauging.  We will not attempt to write down
the detailed form of that gauging, but as T-duality exchanges momentum
and winding modes, the effect should be to kill off all winding modes
while leaving $k$ momentum modes.  This we would like to interpret as
a set of $k$ points.  Thus, in other words, we have a commuting diagram
\begin{displaymath}
\xymatrix@C=2cm{
S^1 \ar[r]^{T-duality} \ar[d]_{ / U(1)} & 
S^1 \ar[d]^{ / dual \: U(1) } \\
[pt/{\bf Z}_k]  \ar[r]^{our \: duality} & \coprod_1^k pt
}
\end{displaymath}
At this same level of handwaving, we can also relate our duality
to T-duality for fibered $S^1$'s as presented in \cite{bem}.
That reference argues that T-duality will exchange $c_1$ of an
$S^1$ bundle with (the pushforward along fibers of) $H$ flux.
For example, if we start with a nontrivial $S^1$ bundle with $H=0$,
then the T-dual will be a trivial $S^1$ bundle with $H \neq 0$.
Given that picture, if we quotient the $S^1$ fibers by $U(1)$ rotations,
then one is led to the picture that ${\bf Z}_k$ gerbes
(which are $B{\bf Z}_k$-bundles) should be T-dual to disconnected
sums of spaces with $B$ field flux, which is certainly part of our
decomposition conjecture.

\section{Applications}    \label{apps}

\subsection{Curve counting predictions}

\subsubsection{Quantum cohomology}

In our previous work \cite{glsm} we observed that Batyrev's conjecture
for the quantum cohomology ring of toric varieties, easily generalizes
to toric stacks.  In particular, Batyrev's conjecture has a precise
physical meaning in terms of the effective action of a UV theory, 
the gauged linear sigma model.
If the toric stack is described in the form
\begin{displaymath}
\left[ \frac{ {\bf C}^N \: - \: E }{ ({\bf C}^{\times})^n } \right]
\end{displaymath}
where $E$ is some exceptional set, and the weight of the $i$th vector
in ${\bf C}^N$ under the $a$th ${\bf C}^{\times}$ is denoted $Q_i^a$,
then Batyrev's conjecture is that the quantum cohomology ring is of the
form ${\bf C}[\sigma_1, \cdots, \sigma_n]$ modulo the relations
\begin{displaymath}
\prod_{i=1}^N \left( \sum_{b=1}^n Q_i^b \sigma_b  \right)^{Q_i^a} \: = \:
q_a
\end{displaymath}
for a set of constants $q_a$.

For example, the quantum cohomology ring of a ${\bf Z}_k$ gerbe over
${\bf P}^N$ with characteristic class $-n \mbox{ mod }k$
(realized as a ${\bf C}^{\times}$ quotient of a principal
${\bf C}^{\times}$ bundle over ${\bf P}^N$)
is given by
\begin{displaymath}
{\bf C}[x,y]/(y^k - q_2, x^{N+1} - y^n q_1)
\end{displaymath}
This is $k$ copies of the quantum cohomology ring of ${\bf P}^N$,
indexed by the value of $y$.

As an aside, note that 
physically, sending the gauge coupling to zero not only 
reduces the quantum cohomology ring to the ordinary cohomology ring,
but also reduces the gerbe to the underlying space:
the distinction between a gerbe and a space is only visible via
nonperturbative effects in physics, so if those nonperturbative 
effects are suppressed, then physics is unable to distinguish the
gerbe from the underlying space.

More generally, a gerbe structure is indicated from the ${\bf C}^{\times}$
quotient description whenever ${\bf C}^{\times}$ charges are nonminimal.
In such a case, from our generalization of Batyrev's conjecture, at least 
one relation in the quantum cohomology ring will have the
form $p^k = q$, where $p$ is a relation in the quantum cohomology of the
underlying toric variety, and $k$ is the greatest common divisor of the
nonminimal charges.  One can rewrite this in the same form as in the
example above of a gerbe on a projective space, and so in this fashion
we can see our decomposition conjecture inside our generalization of
Batyrev's conjecture for quantum cohomology to toric stacks.

\subsubsection{Gromov-Witten predictions}

The fact that the CFT of a string on a gerbe matches the CFT of a disjoint
union of spaces makes a prediction for Gromov-Witten theory:
Gromov-Witten invariants of a gerbe \cite{agv} should be computable in terms of
Gromov-Witten invariants of a disjoint union of spaces.

We shall not give a general proof of that claim for Gromov-Witten
invariants here, but rather will summarize the results of some
calculations in a few special cases.

The first example is $X \times B {\bf Z}_k$, or,
equivalently, $[X/{\bf Z}_k]$ where the ${\bf Z}_k$ acts trivially
globally.  In this case of a trivial ${\bf Z}_k$ gerbe, 
our decomposition conjecture says
\begin{displaymath}
\mbox{CFT}([X/{\bf Z}_k]) \: = \:
\mbox{CFT}\left( \coprod_1^k X \right)
\end{displaymath}
We can understand correlation functions as follows.
Let $\Upsilon$ denote the generator of the $B {\bf Z}_k$ twist
fields, and $f$ a product of twist-field-independent correlators.
In the case of no twist field insertions,
the conjecture predicts
\begin{displaymath}
< f >_{X \times B{\bf Z}_k, g} \: = \: A^{2g-2} k <f>_{X,g}
\end{displaymath}
where $g$ is the genus of the worldsheet and
$A$ is a convention-dependent factor that can be absorbed into
a dilaton shift.
Since the twist fields decouple from the rest of the correlators,
we can write a slightly more general set of correlation functions as
\begin{displaymath}
< f \prod_i \Upsilon^{n_i} >_{X \times B{\bf Z}_k, g} \: = \:
\left\{ \begin{array}{cl}
A^{2g-2} k <f>_{X, g} & \sum_i n_i \equiv 0 \mbox{ mod } k \\
0 & \mbox{else}
\end{array} \right.
\end{displaymath}
If we specialize to $k=2$, then the projectors onto each component of the CFT are given by
\begin{displaymath}
\Pi_1 \: = \: \frac{1}{2} \left( 1 + \Upsilon \right), \: \: \:
\Pi_2 \: = \: \frac{1}{2} \left( 1 - \Upsilon \right)
\end{displaymath}
which obey $\Pi_1^2 = \Pi_1$, $\Pi_2^2 = \Pi_2$, $\Pi_1 \Pi_2 = 0$.
Thus, for example,
\begin{displaymath}
< f \Pi_1 \Pi_2 >_{X \times B {\bf Z}_2, g} \: = \: 0
\end{displaymath}
These results seem consistent with Gromov-Witten calculations
for $B {\bf Z}_k$ \cite{graberpriv}.

Another example is the banded ${\bf Z}_2$ gerbe defined by
$[{\bf C}^3/D_4]$ where the ${\bf Z}_2$ center of $D_4$ acts
trivially.  We discussed this example in section~\ref{firstd4} 
and saw explicitly that the CFT of this noneffective orbifold coincides
with the CFT of the disjoint union of two copies of
$[{\bf C}^3/{\bf Z}_2 \times {\bf Z}_2]$, one with discrete torsion,
the other without.

We would like to thank J.~Bryan for supplying specific Gromov-Witten
calculations in this example.  Let $\{a, b, c\}$ and $\{z, a, b, c\}$
denote nontrivial conjugacy classes in ${\bf Z}_2 \times {\bf Z}_2$ and
$D_4$, respectively, then the basic relationship between genus $g$
invariants of the gerbe and genus $g$ invariants of $[{\bf C}^3/
{\bf Z}_2 \times {\bf Z}_2]$ is
\begin{displaymath}
< z^n a^i b^j c^k >_g \: = \:
\left\{ \begin{array}{cl}
64^{g-1} 2^{1+2(i+j+k)} < a^i b^j c^k >_g & n \mbox{ even} \\
0 & n \mbox{ odd}
\end{array} \right.
\end{displaymath}
In terms of the genus $g$ potential functions $F'_g$ (for the gerbe)
and $F_g$ (for $[{\bf C}^3/{\bf Z}_2 \times {\bf Z}_2]$), one gets the
relation
\begin{displaymath}
F'_g(z, a, b, c) \: = \: 64^{g-1} 2 \cosh(z) F_g(4a, 4b, 4c)
\end{displaymath}

We can see the decomposition conjecture in the results above.
From the decomposition conjecture, a correlation function
on the gerbe should decompose as a sum over correlation functions
in the underlying effective orbifolds.  These particular correlation
functions are insensitive to discrete torsion, and so should be identical
between the two copies.  The projection operators are
$\Pi_{\pm} = \frac{1}{2}(1 \pm z)$, so for example in the notation above
one should find
\begin{displaymath}
< a^i b^j c^k >_{[{\bf C}^3/D_4], g} \: \propto \: 
2 < a^i b^j c^k >_{ [ {\bf C}^3/{\bf Z}_2 \times {\bf Z}_2], g}
\end{displaymath}
up to convention-dependent normalization factors.
The factor of $64^{g-1}$ is a physically-meaningless convention-dependent
dilaton shift (see appendix~\ref{dilatonshift}), and the factors of
$2^{2(i+j+k)}$ can be absorbed into the definition of the vertex operators.
(Those factors almost certainly arise because the computation in the
$D_4$ orbifold was performed with vertex operators realizing sums over
representatives of conjugacy classes rather than averages; in any
event, these also are convention dependent.)
Stripping out convention-dependent factors leaves only a factor of $2$,
exactly as predicted by the decomposition conjecture.

In passing, these ideas also seem consistent with general
aspects of the Gromov-Witten-Donaldson-Thomas correspondence.
After all, we are saying that Gromov-Witten invariants of gerbes
should decompose into Gromov-Witten invariants of spaces;
on the other hand, it is a standard result that sheaves on gerbes
decompose into (twisted) sheaves on spaces, hence Donaldson-Thomas
invariants on gerbes should possess the same decomposition as that
we are claiming for Gromov-Witten invariants of gerbes.

\subsection{Gauged linear sigma model analyses}

In this section we shall analyze a gauged linear sigma model,
in order to show that ``gerby'' effects are common even in descriptions
of ordinary spaces, and so an understanding of gerbes is important
to understand generic gauged linear sigma models for ordinary spaces.

Consider a gauged linear sigma model describing the complete intersection
of four degree-two hypersurfaces in ${\bf P}^7$ at large radius.
This gauged linear sigma model has a total of twelve chiral superfields,
eight ($\phi_i$, $i \in \{ 1, \cdots, 8 \}$) of charge 1
corresponding to
homogeneous coordinates on ${\bf P}^7$, and four ($p_a$,
$a \in \{ 1, \cdots, 4 \}$) of charge $-2$ corresponding 
to the four hypersurfaces.

The D-term for this gauged linear sigma model reads
\begin{displaymath}
\sum_i | \phi_i |^2 \: - \: 2 \sum_a | p_a |^2 \: = \: r
\end{displaymath}

When $r \gg 0$, then we see that not all the $\phi_i$ can vanish,
corresponding to their interpretation as homogeneous coordinates
on ${\bf P}^7$.  More generally, for $r \gg 0$ we recover the
geometric interpretation of this gauged linear sigma model as
a complete intersection.

For $r \ll 0$, we find a different story.
There, the D-term constraint says that not all the $p_a$'s can
vanish; in fact, the $p_a$'s act as homogeneous coordinates on
a ${\bf P}^3$, except that these homogeneous coordinates have
charge 2 rather than charge 1.  That is precisely how one
describes, in gauged linear sigma model language,
the banded ${\bf Z}_2$ gerbe over ${\bf P}^3$ whose characteristic
class is $-1 \mbox{ mod } 2$.

Ordinarily, we would identify this phase with a `hybrid' Landau-Ginzburg
phase, in which one has a family of ordinary Landau-Ginzburg models
fibered over the base.  In this case, however, because the
hypersurfaces are degree two, we get a series of mass terms
for the $\phi_i$, so that at generic points on the gerbe of $p_a$'s,
the $\phi_i$ are all massive and can be integrated out.

If one then works locally in the large $-r$ phase,
since we have a ${\bf Z}_2$ gerbe structure locally,
one might loosely argue that we should really be thinking
in terms of a 2-fold cover of ${\bf P}^3$.

Globally we should be more careful, and take into account the subvariety
where some of the $\phi_i$ become massless.  We can determine that subvariety
as follows.  Write the superpotential in the form $v^T A v$,
where $v = [ \phi_0, \cdots, \phi_7]^T$ is the vector of homogeneous
coordinates on ${\bf P}^7$, and the $8\times 8$ matrix $A$ has entries
that are linear in the $p_a$'s.  Whenever a set of $\phi$'s become
massless, the matrix $A$ will have a vanishing eigenvalue, and so its
determinant will vanish.  Thus, the set of points on ${\bf P}^3$,
the manifold of $p_a$'s, where some $\phi$'s may become massless
is given by $\mbox{det }A$, a degree-eight polynomial.
In particular, thinking in terms of gerbes we should really be considering
a double cover of ${\bf P}^3$ branched over a degree-eight hypersurface
in ${\bf P}^3$.

Now, a double cover of ${\bf P}^3$ branched over a degree 8
hypersurface in ${\bf P}^3$ is an example of a Calabi-Yau.
We can see this as follows.
Let the double cover be denoted $S$, with projection map
$\pi: S \rightarrow {\bf P}^3$, and let a hypersurface class
in ${\bf P}^3$ be denoted $H$.
Assume $S$ is branched over a degree $d$ hypersurface in ${\bf P}^3$.
Then
\begin{displaymath}
K_S \: = \: \pi^* K_{ {\bf P}^3 } \: + \: \tilde{D} \: = \:
\pi^* ( -4 H ) \: + \: \tilde{D} 
\end{displaymath}
where $\tilde{D}$ is the divisor in $S$ over the branch locus,
so $2 \tilde{D} = \pi^* ( d H)$.
Thus,
\begin{displaymath}
2 K_S \: = \: \pi^*( (-8 + d) H )
\end{displaymath}
hence $S$ is Calabi-Yau if $d=8$.
For a closely related discussion in the context of a different
example, see \cite[chapter 4.4, p. 548]{gh}.

We should emphasize, however, that the theory in the large $-r$ phase
of this gauged linear sigma model is not the Calabi-Yau 3-fold $S$,
although it is closely related \cite{simdist}.

Double covers of ${\bf P}^3$ branched over a degree 8 hypersurface in
${\bf P}^3$ are known as octic double solids, and are described
in greater detail in {\t e.g.} \cite{clemens1,cynk1}.

\subsection{Geometric Langlands}

The T-duality between gerbes and disconnected spaces that we have
discussed in this paper also implicitly appears in
\cite{kaped,tonyron}.
Section 7.1 of \cite{kaped} discusses how the moduli space of
$G$-Higgs bundles on a curve $C$ has several components,
and the universal bundle over each component is potentially a twisted
bundle, with twisting determined by an element of
$H^2({\cal M}, Z(G))$ which is described as the obstruction to lifting
a universal $G_{ad}$ bundle to a $G_{s.c.}$-bundle, where $G_{s.c.}$ denoted the simply connected cover of $G$.  
Section 7.2 describes how that twisting enters the physics of the
sigma model on the Higgs moduli space:  for each $e_0$,
corresponding to an irreducible representation of $Z(G)$,
there is at least effectively a flat $B$ field on the corresponding
component determined by the image of the element of $H^2({\cal M},
Z(G))$ under $e_0$.

Using the ideas in this paper, this picture can be equivalently
rewritten as a sigma model on a banded $Z(G)$-gerbe over the moduli space,
whose characteristic class is the element of $H^2({\cal M},Z(G))$
determined in \cite[section 7.1]{kaped}.
As described elsewhere in this paper, a string on a banded $Z(G)$ gerbe
is equivalent to a string on a disjoint union of spaces, one copy
for each irreducible representation of $Z(G)$, with a flat $B$ field
determined by the characteristic class of the gerbe via the map
precisely described in \cite[section 7.2]{kaped}.

In \cite{tonyron}, this part of geometric Langlands was rewritten
in terms of a moduli stack,
which had the structure of a $Z(G)$ gerbe over the moduli space.
We now see that the mathematical manipulations of \cite{tonyron}
can be given a direct explicit physical meaning:  instead of
talking about sigma models on Higgs moduli {\it spaces},
we could just as well work with sigma models on the Higgs
moduli {\it stacks} discussed in \cite{tonyron}.

This gerbiness also has an analogue in the four-dimensional gauge
theory of \cite{kaped}.  The gerbiness appears in two-dimensions
when $Z(G)$ is nontrivial, and corresponds to a four-dimensional gauge
theory in which part of the gauge group is acting ineffectively.
In \cite{kaped}, the matter content of the four-dimensional theory
transforms in the adjoint representation of the group.
If the gauge group is the simply-connected cover $G_{s.c.}$
rather than the centerless adjoint group $G_{ad}$,
then the center acts trivially on the matter, giving a four-dimensional
analogue of the two-dimensional gauge theories we have discussed
here and in \cite{glsm,msx,nr}.  For example, the Higgs moduli space
can now be interpreted as a $Z(G)$-gerbe.  Dimensionally reducing this gerby
picture of the four-dimensional theory also leads directly
to a picture of sigma models on Higgs moduli stacks.

In passing, we should mention that nonperturbative distinctions between 
four-dimensional gauge theories with noneffectively-acting gauge groups
and their effectively acting counterparts have been studied by M. Strassler
in the context of Seiberg duality, see for example 
\cite{strassler1}.

\subsection{Speculations on nonsupersymmetric orbifolds}

In nonsupersymmetric orbifolds, it has been suggested 
\cite{moorenonsusy,evanonsusy} that the RG endpoint
of tachyon condensation involves breaking a connected space into a disjoint
union of spaces.
In this section we will speculate on a physical mechanism by which this
could occur, based on the results of this paper.

The tachyons whose condensation leads to RG flow arise as twist fields,
and are localized at orbifold points.
Now, in terms of stacks, the orbifold ``point'' is replaced by
a gerbe over a point -- this is part of why a quotient stack can be
smooth even when the corresponding quotient space is singular.
Naively, if one imagines a wavefront expanding outward from such a 
singularity describing a dynamical process in which a twisted sector
tachyon gets a vev, then one might imagine that to be consistent
the gerbe structure should be carried along, expanding to cover
the entire space.
Perhaps the endpoint of tachyon condensation and RG flow should
be understood in terms of a gerbe structure covering the original
space.  From the observations of this paper, a string on a gerbe
is equivalent to a disjoint union of spaces, so perhaps the
suggestion of \cite{moorenonsusy,evanonsusy} that disconnected spaces
arise at the endpoint is just the T-dual picture.

At this point we should mention the analogous idea does not work
in supersymmetric cases.  For example, consider the quotient stack
$[{\bf C}^2/{\bf Z}_2]$ where the ${\bf Z}_2$ acts by sign flips on 
the coordinates of ${\bf C}^2$.  The corresponding quotient space has
a singularity at the origin, whereas the stack has a gerbe at the origin.
Naive blowups of this stack at the origin result in stacks over
$\widetilde{ {\bf C}^2/{\bf Z}_2 }$ with a ${\bf Z}_2$ gerbe
over the exceptional divisor.  It was originally suggested in
\cite{meold} that this could be a mechanism for understanding old
lore concerning ``$B$ fields at orbifold points'' \cite{edstr95,paul95}.
After all, if a string on a gerbe is related to strings on underlying
spaces with $B$ fields, then perhaps a string on a stack that looks
like a ${\bf Z}_2$ gerbe over the exceptional divisor should have the
same CFT as a string on the underlying space with a flat $B$ field.
This original proposal, unfortunately, suffers from the problem that the
stack in question, the blowup of $[ {\bf C}^2/{\bf Z}_2]$,
is not Calabi-Yau, and so should not arise when deforming along flat
directions in the supersymmetric theory.

In the nonsupersymmetric case, the constraint on supersymmetric flat
directions is weakened somewhat, so perhaps some version of this
picture should apply.

We will not discuss this further here, but thought it appropriate to mention
that this is a natural speculation considering the topic of this paper.

\section{Conclusions}    \label{concl}

In this paper we have described T-duality between strings propagating on 
gerbes and strings propagating on disconnected
sums of spaces.  This duality solves a basic problem with the
notion of strings propagating on gerbes, namely that the massless spectrum
violates cluster decomposition, one of the foundational axioms of quantum
field theory.  A sigma model on a disjoint union of spaces also violates
cluster decomposition, but in the mildest possible way, hence a T-duality
between gerbes and disjoint unions of spaces means that the CFT's associated
to gerbes must also be consistent, despite violating cluster decomposition.

We presented a detailed description of how gerbes should decompose.
The disjoint union will, in general, consist of different spaces, not a 
sum of copies of the same space, and there will be different $B$ fields
on each component.  After presenting a conjecture valid for all cases,
we began giving evidence that this conjecture was correct.
In examples presented as global quotients by finite groups, we 
computed partition functions (at arbitrary genus) and massless spectra
in a number of examples, and checked that this conjecture held true
in each case.  For gerbes presented as global quotients by nonfinite groups,
we also discussed how this conjecture could be seen directly in the structure
of nonperturbative effects in the theories.
We also discussed open strings in this context.  Because of the relation 
between D-branes and K-theory, this conjecture makes a prediction for
equivariant K-theory, which we checked in appendix \ref{mandoproof}.  
For D-branes in the open string B model, which can be described
by sheaves and derived categories thereof, our decomposition conjecture
corresponds to a known decomposition result for sheaves on gerbes.
We also discussed how other aspects of open strings in the B model
are consistent with this conjecture.
We discussed mirror symmetry in this context, showing how the
results on mirror symmetry for stacks presented in \cite{glsm}
are consistent with this decomposition conjecture, and also
discussed a noncommutative-geometry-based argument for this conjecture.
We discussed the effect of discrete torsion along noneffective directions,
and also discussed why this decomposition can be understood as a T-duality.

Finally, we discussed applications of these ideas.
One important set of applications is to curve-counting; in effect,
we are making predictions for quantum cohomology and, more generally,
Gromov-Witten invariants.  In previous work \cite{glsm} we presented
an analogue of Batyrev's conjecture for quantum cohomology of toric stacks,
and the decomposition conjecture can be seen in that conjecture.
We also discuss how the predictions for Gromov-Witten theory work out
in a few simple examples.

We also discuss how these ideas are relevant to analyses of ordinary
gauged linear sigma models, and in fact their appearance there
suggests that these ideas are much more widely relevant to gauged
linear sigma models than one might have naively supposed.

We also discuss how these ideas can be used to give a minor improvement
in the physical understanding of the geometric Langlands program,
by giving a direct physical relationship between different
descriptions of geometric Langlands in terms of disconnected spaces
and in terms of gerbes.

Finally, we speculate on how these ideas may also give an alternative
understanding of some aspects of tachyon condensation in nonsupersymmetric
orbifolds.

\section{Acknowledgements}

We have discussed these matters with a number of people.
In particular, we would like to mention
J.~Bryan, R.~Cavalieri, J.~Distler, T.~Graber,
T.~Jarvis, S.~Katz, C.~Phillips, R.~Plesser, M.~Rieffel,
K.~Wendland, and E.~Witten.
The work of S.H. was supported in part by the DOE under grant
DE-FG02-90ER40542.  S.H. is
the D.E. Shaw \& Co., L. P. Member at the Institute for
Advanced Study.
T.P. was supported in part by NSF grant DMS-0403884
and NSF FRG grant DMS-0139799.
E.S. would like to thank the math/physics research group at the
University of Pennsylvania for hospitality while much of this work
was conducted.
M.A. was supported in part by NSF grant DMS-0306429.

\appendix

\section{Partition functions for disconnected targets}  \label{dilatonshift}
 
What is the partition function for a nonlinear sigma model
on the disjoint union of $k$ copies of a manifold $X$?
                                                                                
There are two clear possible answers, and strong arguments supporting both.
                                                                                
One possibility is that at any worldsheet genus,
the partition function for $\coprod_k X$ should be $k$ times the partition
function for $X$.  We can justify this by thinking about the path
integral measure:  a sum over maps from a connected worldsheet
into a disconnected target can be decomposed into a sum over
target components, with a sum over maps into each component.
Thus, at any genus $g$, the partition function
for a nonlinear sigma model on $\coprod_i X_i$ should have the form
\begin{displaymath}
Z_g(\coprod_i X_i ) \: = \: \sum_i Z_g(X_i)
\end{displaymath}
and so, in particular, the genus $g$ partition function
for a nonlinear sigma model on the disjoint union of $k$ copies
of the same space $X$ should be given by
\begin{displaymath}
Z_g(\coprod_i X) \: = \: k Z_g(X)
\end{displaymath}
                                                                                
An alternative argument gives a different result.
Consider the disjoint union of two copies of $X$.
If we orbifold the nonlinear sigma model with that target
by a ${\bf Z}_2$ which exchanges the two copies,
since the group acts freely, the orbifold must be the nonlinear
sigma model whose target is one copy of $X$.
At genus $g$, the partition function of the orbifold theory
is easily computed to be
\begin{displaymath}
Z_g({\bf Z}_2) \: = \: \frac{1}{| {\bf Z}_2 |^g} \sum_{t.s.} Z_{t.s.}(X
\sqcup X)
\end{displaymath}
where the $t.s.$ denotes the twisted sector boundary conditions that
we must sum over.  Now, since the ${\bf Z}_2$ exchanges the two
copies of $X$, and the worldsheet is connected,
any $Z_{t.s.}$ with a nontrivial element of ${\bf Z}_2$ at any
boundary must vanish, as there are no continuous
maps from a connected worldsheet
that touch both copies of $X$ in the target.
Thus,
\begin{displaymath}
\sum_{t.s.} Z_{t.s.}(X \sqcup X) \: = \: Z_g(X \sqcup X)
\end{displaymath}
and so we recover the algebraic identity that
\begin{displaymath}
Z_g(X \sqcup X) \: = \: 2^g Z_g(X)
\end{displaymath}
This result does agree with the intuition that on a genus $g$
worldsheet, since the nonlinear sigma model on the disjoint
union of $k$ copies of $X$ has $k$ times as many states as
a nonlinear sigma model on $X$, the genus $g$ partition function
should have a factor of $k^g$, reflecting the fact that
one expects a sum over states along each handle of the worldsheet.

Unfortunately, these two perspectives conflict -- these two
ways of thinking about the partition function on a disjoint union
of copies of $X$ are giving us different results.
The first argument said that the numerical factor should be
independent of genus, the second said it is not independent of
genus.  Only at genus one do the two arguments give the same result.
                                                                                
Now, since the partition functions only
differ by a numerical factor, the reader might think this is
not important.
After all, in ordinary QFT such numerical factors are irrelevant.
However, worldsheet string theory is a theory coupled to worldsheet gravity,
and in a theory coupled to gravity, factors in front of
partition functions are important, and reflect state degeneracies,
as discussed in more detail in \cite{nr}.
We cannot ignore differing numerical factors, as that would be
tantamount to ignoring the cosmological constant.
                                                                                
The resolution of this puzzle is that these two ways of
thinking about the genus $g$ partition function of a theory
on a disjoint union differ by a dilaton shift.
If we shift the vev of the dilaton by a constant, say, $A$,
then the $g$-loop partition function shifts by a factor as
\begin{displaymath}
Z_g \: \mapsto \: (\exp(A))^{2g-2} Z_g
\end{displaymath}
since in the two-dimensional action the dilaton multiplies the
worldsheet Ricci scalar, and so shifting the dilaton by a constant
will multiply the path integral by a factor raised to the power
of the Euler characteristic of the worldsheet.

In any event, such dilaton shifts are physically trivial,
and so any two partition functions differing by factors
raised to the $2g-2$ power are physically equivalent.

\section{Miscellaneous group theory}

This paper uses an unusually large amount of group theory which may
not be at the reader's fingertips, so for reference
in this appendix we summarize a few relevant standard finite group theory 
results.

The number of elements of a group is the sum of the squares of the
dimensions of the irreducible representations \cite[eqn'n (1.74)]{georgi}:
\begin{displaymath}
|G| \: = \: \sum_i ( \mbox{dim } \rho_i )^2
\end{displaymath}
An irreducible representation $\rho$ appears in the regular representation
$\mbox{dim }\rho$ times \cite[cor 2.18]{fh}.

The number of irreducible representations of $G$ is equal to the number
of conjugacy classes of $G$ \cite[prop. 2.30]{fh}.

For example, consider the eight-element dihedral group $D_4$.
The elements of this group are
\begin{displaymath}
D_4 \: = \: \{ 1, z, a, b, az, bz, ab, ba=abz\}
\end{displaymath}
where $z$ generates the center, $z^2=1=a^2$, $b^2=z$.
This group has five conjugacy classes, given by
\begin{displaymath}
1, z, \{a, az\}, \{ b, bz\}, \{ab, ba\}
\end{displaymath}
First, let us count the one-dimensional representations of $D_4$.
These are homomorphisms $f: D_4 \rightarrow {\bf C}^{\times}$.
Thus, $f(1) = 1$, and since the projection $D_4 \rightarrow {\bf Z}_2 \times
{\bf Z}_2$, $f(a) = f(az)$, hence $f(z)=1$.
Define
\begin{eqnarray*}
\alpha & = & f(a) \\
\beta & = & f(b) 
\end{eqnarray*}
then $\alpha^2 = \beta^2 = 1$, and so we have four possible
one-dimensional representations, indexed by the values of $\alpha$ and
$\beta$.
In addition, there is a two-dimensional representation given by
\begin{displaymath}
z \: = \: \left[ \begin{array}{cc}
                 -1 & 0 \\
                 0 & -1 \end{array} \right], \: \:
a \: = \: \left[ \begin{array}{cc}
                 0 & 1 \\
                 1 & 0 \end{array} \right], \: \:
b \: = \: \left[ \begin{array}{cc}
                 i & 0 \\
                 0 & -i \end{array} \right]
\end{displaymath}

Since there are 5 conjugacy classes, this must be all of the
irreducible representations.
Also, note that the sum of the squares of the dimensions of the
irreducible representations equals the order of $D_4$.

\section{Group extensions and twisted K-theory (written by M.~Ando)}   
\label{mandoproof}

The conjecture in the paper suggests the following
result about $K$-theory.  Let 
\[
   1 \rightarrow G \rightarrow H \rightarrow  Q \rightarrow 1
\]
be an extension of compact Lie groups.  Let $X$ be a $Q$-space.
Then $K_{H} (X)$ is isomorphic to the twisted $Q$-equivariant
$K$-theory of $X$, where the twist is the one described in the main
body of the paper (and below).

This problem has been extensively studied, particularly from the
$C^{*}$-algebra point of view; see for example \cite{MR1292014}.  M.A. is 
grateful to Chris Phillips and Marc Rieffel for
pointing this out, and for several very helpful conversations.

In this appendix we sketch a proof of
this result, at least when the groups are finite, and the extension is
a semi-direct product.  Details and generalizations will appear later.

M.A. first learned some of the ideas used here in the course of
joint work in preparation with John Greenlees.

\subsection{The category of irreducible representations}

Let $G$ be a compact Lie group.  Let $\mbox{Irr}(G)$ be the category of
irreducible complex representations of $G$: an object of $\mbox{Irr}(G)$ is
an irreducible representation 
\[
   \alpha:  \: G\: \rightarrow \: \mbox{GL}(V),
\] 
where $V$ is a complex vector space.  A morphism is an isomorphism  
\[
    f: \: V\: \rightarrow \: V'
\]
which intertwines the action of $G$ on $V$ and $V'$.  This category is
a groupoid.  By Schur's Lemma, for every object $\alpha: G\rightarrow 
\mbox{GL} (V)$
of $\mbox{Irr}(G)$, the canonical map 
\[
   {\bf C}^{\times}\: \rightarrow \:  \mbox{Aut}(V)
\]
is an isomorphism. (Also, it commutes with any morphism $f$) In
practice, it is useful to set this up so that 
the objects of $\mbox{Irr}(G)$ comprise a set: 
we can and will do this by choosing a
set $G^{\vee}$ of representatives for the irreducible representations
of $\pi_{0}\mbox{Irr}(G)$.  

When it is convenient, we shall write $V_{\alpha}$ for the vector
space associated to a representation $\alpha$.

\subsection{Irreducible representations and equivariant vector bundles}
\label{sec:irred-repr-equiv}

For example, if $V$ is a
$G$-equivariant vector bundle over a $G$-fixed space, then we have a
canonical map of equivariant vector bundles
\begin{equation}\label{eq:1}
      \bigoplus_{\alpha\in G^{\vee}} \alpha\otimes \hom (\alpha,V)
      \: \stackrel{\cong}{\longrightarrow} \:  V,
\end{equation}
which is an isomorphism by Schur's Lemma.   On the left, $\alpha$ is
pulled back along 
\[
       X \: \rightarrow \: \mbox{pt},
\]
while $G$ acts trivially on $\hom(\alpha,V).$   The same result, with
the same formula, applies whether $V$ is a genuine or virtual $G$-vector bundle.

\subsection{Group extensions and equivariant projective bundles}
\label{sec:group-extens-equiv}

Suppose that, as in \S\ref{sec:irred-repr-equiv}, we have a $G$-vector
bundle or $K$-theory object $V$ over a $G$-fixed space $X$.   
In addition we suppose that $G$ participates in a short exact sequence
\begin{equation}\label{eq:4}
   1 \: \longrightarrow \: G \: \longrightarrow \: H \: \longrightarrow
\:  Q \: \longrightarrow \: 1
\end{equation}
of compact Lie groups, and $Q$ acts on $X$. (We use $Q$ rather
than $K$ as in the main text to denote the quotient, to avoid
talking about $K$-equivariant $K$-theory)

In this situation we can ask whether $H$ 
can be made to act on $V$, in such a way that the restriction of the
action to $G$ is the given one; indeed we can ask for the space of
solutions to this problem.  It's simplest to analyze the situation in
the case that all the groups in sight are finite, and $H$ is a
semidirect product.
We treat that case.

Let $k\in Q$.  Pulling back the
isomorphism (\ref{eq:1}) along 
\[
k: \: X\: \rightarrow \: X
\]
gives 
\begin{equation}\label{eq:2}
      \bigoplus_{\alpha\in G^{\vee}} k^{*}\alpha\otimes 
k^{*}\hom (\alpha,V)
      \: \stackrel{\cong}{\longrightarrow} \: k^{*}V.
\end{equation}
To make $Q$ act on $V$ compatibly with its $G$-action, we must give
an isomorphism of $G$-objects over $X$
\[
  \phi_{k}: \: 
\bigoplus_{\alpha\in G^{\vee}}  \alpha\otimes \hom (\alpha,V)
\: \stackrel{\cong}{\longrightarrow} \:
  \bigoplus_{\alpha\in G^{\vee}} k^{*}\alpha\otimes k^{*}\hom
  (\alpha,V).
\]
As $k$ varies these isomorphism must satisfy the cocycle condition 
\[
       \ell^{*}\phi_{k} \phi_{\ell} = \phi_{k\ell}.
\]
The subtlety arises from the fact that $k^{*}\alpha$ is isomorphic to
an element of $G^{\vee}$, but not canonically. 

For any representation
\[
\rho  : G\to \mbox{GL} (V),
\]
let $k^{*}\rho$ be the representation 
\[
      G \: \stackrel{c_{k}}{\longrightarrow}\:
 G \: \stackrel{\alpha}{\longrightarrow} \: \mbox{GL} (V),
\]
where $c_{k}$ is conjugation by $Q$ in the semidirect product.
For concreteness, let
\[
     k^{*}\rho (g) \: = \: \rho(k^{-1}g k),
\]
so $Q$ acts on the left: 
\[
    (k\ell)^{*}\rho = k^{*} (\ell^{*}\rho).
\]

Clearly $\rho$ is an irreducible representation of $G$ if and only if
$k^{*}\rho$ is, so if $\alpha\in G^{\vee}$, then $k^{*}\alpha$ is 
isomorphic to an element $k (\alpha)\in G^{\vee}$.  Thus the action of
$Q$ on $G$ determines an action of $Q$ on the left of $G^{\vee}.$
Note that these data \emph{do not} determine an isomorphism  
\[
     \beta: \: k^{*}\alpha \: \cong \: k (\alpha).
\]
However, any two isomorphisms $\beta$, $\beta'$ differ by an element of 
\[
       \mbox{Aut} (k (\alpha))\: \cong \:  {\bf C}^{\times}.
\]

{\bf Definition}.  
Let 
\[
{\cal W} \: \longrightarrow \: G^{\vee}
\]
be the bundle of projective spaces over $G^{\vee}$ whose fiber at
the representation 
\[
     \alpha: \: G \: \longrightarrow \: \mbox{GL} (V_{\alpha})
\]
is ${\bf P} (V_{\alpha})$. 
The extension (\ref{eq:4}) gives to ${\cal W}$ the structure of an
$Q$-equivariant bundle of projective spaces over $G^{\vee}$.

It is illuminating to give a cohomological description of the bundle 
${\cal W}$.  
For each $k\in Q$ and $\alpha\in G^{\vee}$,  let's \emph{choose} an
isomorphism  
\[
   \beta (k,\alpha): \: k^{*}\alpha \: \longrightarrow \: k (\alpha).
\]
That is, if $\alpha$ and $k (\alpha)$ are homomorphisms 
\[
    \alpha: \: G\: \longrightarrow \:\mbox{GL} (V)
\]
and 
\[
    k (\alpha): \: G\: \longrightarrow \:  \mbox{GL} (W),
\]
then $\beta (k,\alpha)$ is just vector a space isomorphism 
\[
    \beta (k,\alpha): \: V \: \longrightarrow \:  W
\]
such that 
\[
   \beta (k,\alpha) (\alpha (k^{-1}g k)v)  \: = \:  k (\alpha)\beta
   (k,\alpha) (v).
\]

We then have \emph{two} isomorphisms 
\[
    (k\ell)^{*}\alpha \: \cong \: k\ell (\alpha),
\]
namely 
\[
    \beta (k\ell,\alpha)
\]
and 
\[
    \beta (k,\ell (\alpha)) k^{*}\beta (\ell,\alpha).
\]
Here $k^{*}\beta (\ell,\alpha)$ is the same map of vector spaces as
$\beta (\ell,\alpha),$ but now intertwining $k^{*}\ell^{*}\alpha$ and
$k^{*}\ell (\alpha).$  

Their ratio
\[
    w (k,\ell,\alpha) \: \equiv \: \beta (k,\ell (\alpha))k^{*} \beta
    (\ell,\alpha) \beta (k\ell,\alpha)^{-1}
\]
is an element of $\mbox{Aut} (\alpha)\: \cong \: {\bf C}^{\times}.$

{\bf Lemma}.   
The function 
\[
   w: \: Q^{2}\times G^{\vee} \: \longrightarrow \: {\bf C}^{\times}
\]
is a two-cocycle for the action of $Q$ on $G^{\vee}$.  Its cohomology
class in 
\[
    H^{2}_{Q} (G^{\vee};{\bf C}^{\times}) \: = \:
    H^{2} (EQ\times_{Q}G^{\vee};{\bf C}^{\times}) \: \longrightarrow \:
 H^{3}_{Q}
    (G^{\vee};{\bf Z})  
\]
depends only on the action of $Q$ on $G$.

{\bf Remark}.  We recall how to calculate this sort of Borel
cohomology in \S\ref{sec:borel-cohomology}.

{\it Proof}.
The two-cocycle condition is 
\[
    w (k,\ell,\alpha)w (jk,\ell,\alpha)^{-1}w (j,k\ell,\alpha)
    w (j,k,\ell (\alpha))^{-1} \: = \: 1,
\]
which is easily checked.
  The dependence of $w$ on $\beta$ is as follows.  Any two choices of
isomorphisms $\beta$ and $\beta'$  differ by 
\[
   \beta' (k,\alpha)  \: = \: \delta (k,\alpha) \beta (k,\alpha),
\]
where $\delta (k,\alpha)\in {\bf C}^{\times}\: \cong \: 
\mbox{Aut} (k (\alpha))$.  
This is a function 
\[
    \delta: Q\times G^{\vee}\: \longrightarrow \: {\bf C}^{\times},
\]
and it is easy to see that then 
\[
w'  \: = \: w d\delta,
\]
so $w'$ and $w$ are cohomologous.  $\Box$

\subsection{Equivariant $K$-theory}

For any compact Lie group $H$, equivariant complex $K$-theory
has a classifying space ${\bf K}_H$.  (We also write ${\bf K}$ for a space
representing non-equivariant complex $K$-theory.)
Atiyah and Segal give a number of models for ${\bf K}$ in
\cite{math.KT/0407054}.  For example, one can stabilize the
Grassmannian of finite-dimensional subspaces of a Hilbert
space representation ${\cal U}$ of $H$, in which each irreducible
representation occurs with infinite multiplicity.

In our situation, if $G$ acts trivially on $X$, then clearly any $H$-map 
\[
   X\: \longrightarrow \: {\bf K}_H
\]
factors through 
\[
   {\bf K}_H^{G} \: \longrightarrow \: {\bf K}_H,
\]
and indeed
\[
    \mbox{map}_{H} (X,{\bf K}_H)  \: = \: \mbox{map}_{Q} (X,{\bf K}_H^{G}).
\]
Thus to understand $H$-equivariant $K$-theory of $G$-fixed spaces
amounts to understanding the space ${\bf K}_{H}^{G}$, with its $Q$-action.

The first point is that, if we remember only the $G$-action on
${\bf K}_H$, then ${\bf K}_H$ is a representing space for $G$-equivariant
$K$-theory.   It follows that, if we forget the $Q$-action, then
${\bf K}_H^{G}$ represents $G$-equivariant $K$-theory over $G$-fixed
spaces. The decomposition (\ref{eq:1}) using Schur's Lemma then implies the 
following.

{\bf Proposition}
\begin{equation} \label{eq:3}
      {\bf K}_H^{G} \: \simeq \: \mbox{map} (G^{\vee},{\bf K}).
\end{equation}
$\Box$

{\bf Remark}.   
Of course this is just a formulation of the isomorphism 
\[
    K_{G} (X) \: \cong \: R (G)\otimes K (X)
\]
for $G$-fixed spaces $X$ \cite{Segal:EquivariantK}.

We have written $\simeq$ to indicate homotopy equivalence, but in
practice the decomposition will arise from a geometric
decomposition of our chosen representing space for $K_{H}$.   For
example, the Hilbert space ${\cal U}$ will decompose according to
the irreducible representations of $G$.  The right-hand side of (\ref{eq:3})
 can
 profitably be thought of as
the sections of a fiber bundle ${\bf K}_{\bullet}$ over $G^{\vee}$.  The fiber
over $\alpha$ is a space ${\bf K}_{\alpha}$ with a stable vector bundle
$\xi_{\alpha}$ over it.  The group $G$ acts on $\xi_{\alpha}$ in such
a way that the natural map 
\[
     \alpha\otimes \hom (\alpha,\xi_{\alpha})  \: \longrightarrow \: 
\xi_{\alpha}
\]
is an isomorphism.  If we also forget the $G$-action, then
$\xi_{\alpha}/{\bf K}_{\alpha}$ represents nonequivariant complex
$K$-theory.  The problem remains of describing the $Q$-action on
${\bf K}_H^{G}$.    

Atiyah and Segal \cite{math.KT/0407054} show that $H^{3}_{Q} (X;{\bf Z})$
classifies $Q$-equivariant \emph{stable} projective bundles over $X$:
these are bundles $P$ for which $P\cong P\otimes L^{2} (Q).$   To
such a bundle $P$ they attach a $Q$-equivariant bundle ${\bf K} (P)$.  Over
each point $x\in X$, the fiber ${\bf K} (P)_{x}$ is a representing space
for $K$-theory, but the $Q$-action is twisted from the usual
representing space ${\bf K}_{Q}$ by the equivariant projective bundle $P$.  
They define
\[
   K_{Q,P}^{0} (X) = \pi_{0}\Gamma (X,{\bf K} (P))^{Q},
\]
the homotopy classes of $Q$-equivariant sections of $\mbox{K} (P).$

In particular we can form ${\bf K} (w),$ the equivariant $Q$-bundle over
$G^{\vee}$ associated to the two-cocycle $w$.  We call this ${\bf K}
({\cal W})$, because ultimately it comes from this bundle.  The discussion
in \S\ref{sec:group-extens-equiv}  
(and in the main text of the paper) shows that

{\bf Proposition}.  
As $Q$-equivariant bundles over $G^{\vee}$,
\[
  {\bf K}_{\bullet} \: \cong \: {\bf K} ({\cal W}),
\] 
and so 
\[
   \Gamma (G^{\vee},{\bf K} ({\cal W}))
\]
represents the $Q$-space ${\bf K}_H^{G}.$

Thus if $G$ acts trivially on $X$, 
\begin{equation}\label{eq:6}
    K_{H}(X) \: \cong \: [X,\Gamma (G^{\vee},{\bf K} ({\cal W}))]^{Q}.
\end{equation}
The usual adjunctions based on the formula 
\[
    \mbox{map} (X\times Y,Z) \: \cong \: \mbox{map} (X,\mbox{map} (Y,Z)),
\]
show that 
\begin{equation}\label{eq:5}
    [X,\Gamma (G^{\vee},{\bf K} ({\cal W}))]^{Q}
\: \cong \: \pi_{0}\Gamma (X\times G^{\vee},{\bf K} ({\cal W}))^{Q}.
\end{equation}
The right-hand side is just the twisted $Q$-equivariant $K$-theory of
$X\times G^{\vee}$, where the twist is the one obtained by pulling
back ${\cal W}$ along 
\[
   X\times G^{\vee} \: \longrightarrow \: G^{\vee}.
\]
Putting together (\ref{eq:6}) and (\ref{eq:5}) gives
\[
      K_{H} (X) \: \cong \: K_{Q,{\cal W}}(X),
\]
which is is the result indicated in the main body of the paper.

\subsection{Borel cohomology}
\label{sec:borel-cohomology}

Let $Z$ be a space with an action of a topological group $Q$.  Then we can form
the simplicial space ${\cal X}_{\bullet}$ whose $n$ space for $n\geq 0$ is 
\[
     {\cal X}_{n} \: = \: Q^{n}\times Z
\]
and whose face maps 
\[
    d_{i} : \: {\cal X}_{n+1} \: \longrightarrow \: {\cal X}_{n} 
\]
are given by 
\begin{displaymath}
\begin{array}{rclc}   
d_{0} (k_{0}, \cdots , k_{n},z) &= & (k_{1},\cdots ,k_{n},z) & $\,$\\
d_{i} (k_{0}, \cdots, k_{n},z) &= &
(k_{0}, \cdots, k_{i-1}k_{i}, \cdots, k_{n},z) &
 1 \leq i \leq n \\
d_{n+1} (k_{0}, \cdots, k_{n},z) &= &
(k_{0}, \cdots, k_{n-1},k_{n}z). & $\,$
\end{array}
\end{displaymath}
The geometric realization of ${\cal X}_{\bullet}$ is the Borel consruction, 
\[
       |{\cal X}_{\bullet}| = EQ\times_{Q}Z.
\]
This gives a variety of spectral sequences for Borel cohomology.   For
example, if $Q$ and $Z$ are discrete, then we find that 
\[
         H^{2} (EQ\times_{Q}Z;{\bf C}^{\times})
\]
is described by precisely the sorts of cocycles modulo coboundaries we
use in the lemma above. 

\end{document}